\documentclass[12pt]{article}
\usepackage{a4wide,epsfig}
\usepackage{scalefnt}

\usepackage[latin1]{inputenc}
\usepackage{amsmath,amssymb}
\usepackage{amsfonts}
\usepackage{amssymb}
\usepackage[english]{babel}
\usepackage{graphicx}
\usepackage[caption=false]{subfig}
\usepackage{slashed}

 \usepackage{cite}
\begin{document}

\title{
\begin{flushright}
\small TTP12-048\\
SFB/CPP-12-103\\
\end{flushright}
\vskip 1.5cm
Penguin diagrams in the charm sector in $K^+\to \pi^+\nu\bar{\nu}$}
\author{J.~Mond\'ejar and J.~Rittinger
\\[0.5cm]
{\small\it Institut f\"ur Theoretische Teilchenphysik, Karlsruhe Institute of Technology (KIT),}\\
{\small\it Wolfgang-Gaede-Stra\ss e 1, 726128 Karlsruhe, Germany}\\
}
\date{}

\maketitle

\begin{abstract}
\noindent
We evaluate at next-to-next-to-leading order (NNLO) the QCD corrections to the charm contribution from penguin diagrams to the decay $K^+\to \pi^+\nu\bar{\nu}$. A NNLO calculation is already available in the literature\cite{Buras:2006gb}. We provide an independent check of the results of non-anomalous and anomalous diagrams. We use Renormalization Group improvement and an effective theory framework to resum the large logarithms that appear. In the case of the non-anomalous diagrams, our results for the decoupling coefficients and anomalous dimensions, as well as the final numerical result, are in agreement with those of Ref.~\cite{Buras:2006gb}. In the anomalous case, analytical and numerical disagreements are observed.
\end{abstract}

\section{Introduction}

The rare decay mode $K^+\to \pi^+\nu\bar{\nu}$, along with $K_L\to \pi^0\nu\bar{\nu}$, plays an important role in flavor physics. It probes the quantum  structure of flavor dynamics in the Standard Model (SM) \cite{Vainshtein:1976eu,Ellis:1982ve,Dib:1989cc,Buchalla:1993bv,Misiak:1999yg,Buchalla:1993wq,Buchalla:1998ba,Buras:2006gb} or its extensions \cite{Buras:2003jf,Buras:2005xt,Buras:2004sc,Buras:2000dm,D'Ambrosio:2001zh} while remaining theoretically clean. A recent review can be found in Ref.~\cite{Buras:2004uu}. The cleanness of this decay is the main reason behind its importance, and it is due to the following:
\begin{itemize}
\item{It being a semi-leptonic process, the relevant hadronic operator is just a current operator whose matrix element can be extracted from the leading decay $K^+\to\pi^0 e^+\nu$, including isospin-breaking corrections \cite{Marciano:1996wy}.}
\item{It is short-distance dominated. Long-distance contributions turn out to be small \cite{Isidori:2005xm}, and in principle calculable by means of lattice QCD \cite{Isidori:2005tv}.}
\end{itemize}
As a consequence, this decay can be reliably computed with available field theoretical methods.
Thus, the SM decay rate $K^+\to \pi^+\nu\bar{\nu}$, alone or together with $K_L\to \pi^0\nu\bar{\nu}$, allows for clean determinations of Cabibbo-Kobayashi-Maskawa (CKM) matrix elements. These decays are also very suitable for the search for new physics: a comparison of the determinations of various parameters from studies of  CP violation in B decays with the determinations from the $K\to\nu\bar\nu$ decays 
would provide with a clean signal of physics beyond the SM if substantial deviations were found.

\begin{figure}
\centering
\parbox{40mm}{\includegraphics[width=0.25\columnwidth]{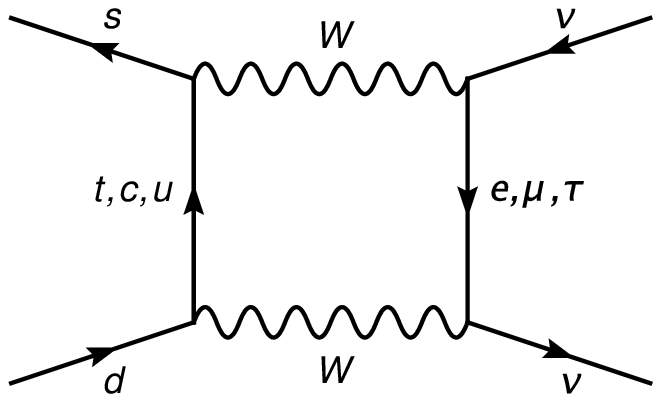}}\qquad\quad \parbox{40mm}{\includegraphics[width=0.25\columnwidth]{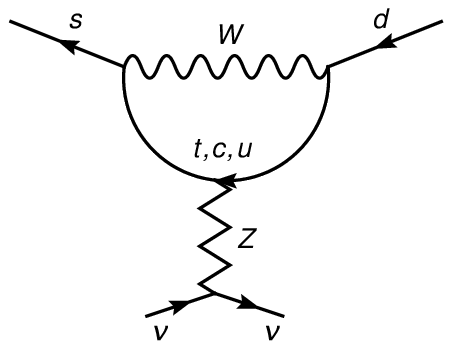}}
\caption{\label{boxandpengdiags}
Box- and penguin-type contributions to $K^+\to \pi^+\nu\bar{\nu}$.}
\end{figure}

\begin{figure}
\centering
\parbox{40mm}{\includegraphics[width=0.28\columnwidth]{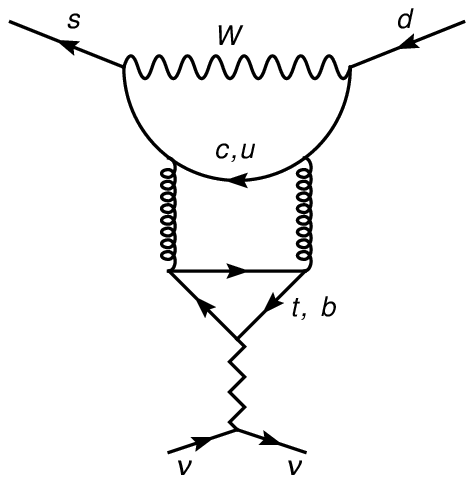}}\qquad\quad \parbox{40mm}{\includegraphics[width=0.28\columnwidth]{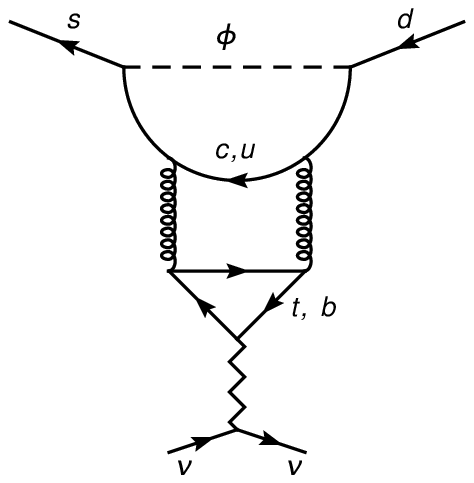}}
\caption{\label{anomdiags}
Two anomalous penguin contributions to $K^+\to \pi^+\nu\bar{\nu}$.}
\end{figure}

\begin{figure}
\centering
\parbox{40mm}{\includegraphics[width=0.22\columnwidth]{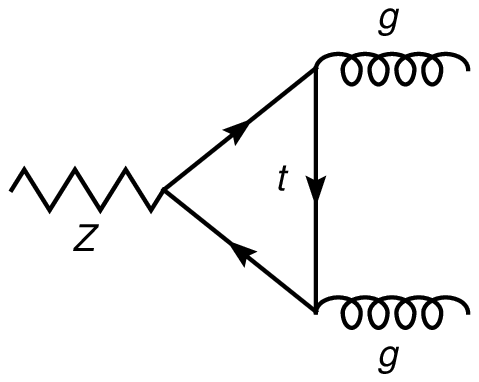}}\qquad\quad \parbox{40mm}{\includegraphics[width=0.22\columnwidth]{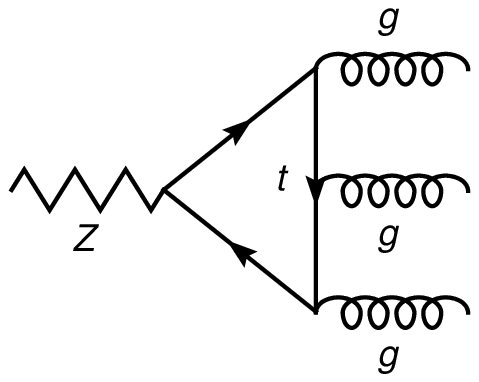}}
\caption{\label{toptrdiags}
The one-loop top quark contribution to the coupling of the $Z$ boson to two and three gluons. Diagrams obtained by the interchange of the external gluons are not shown.}
\end{figure}

In the SM, the $K^+\to \pi^+\nu\bar{\nu}$ decay is mediated by box- and penguin-type diagrams, shown in Figure~\ref{boxandpengdiags}.
The leading, low-energy hamiltonian for this process can be written as \cite{Buchalla:1993wq,Buchalla:1998ba}
\begin{align}
\label{heff}
\mathcal{H}_{\rm{eff}}=\frac{G_F}{\sqrt{2}}\frac{\alpha}{2\pi\,\sin^2\theta_W}\sum_{l=e,\mu,\tau}\sum_{q=t,c,u}V^*_{qs}V_{qd}\,X (x_q)\, \bar{s}\gamma^{\mu}(1 - \gamma_5)d \otimes \bar{\nu}_l\gamma_{\mu}(1 - \gamma_5)\nu_l\,.
\end{align}
Here $G_F$, $\alpha$, and $\sin^2 \theta_W$ denote the Fermi coupling, the electromagnetic coupling, and the weak mixing angle, respectively. $V_{ij}$ are CKM matrix elements and $x_q=m_q^2/M_W^2$.
The contributions mediated by the exchange of virtual top quarks, encoded in the function $X(x_t)$, can be calculated entirely within ordinary perturbation theory. This function is known through next-to-leading order (NLO) \cite{Buchalla:1993bv,Misiak:1999yg}, which offers sufficient accuracy.
The contributions involving only the light $u$, $d$, and $s$ quarks were computed in Ref.~\cite{Isidori:2005xm} using Chiral Perturbation Theory, along with those of subleading, dimension-8 operators.
When considering the charm quark contributions given by $X(x_c)$, one takes the $u$, $d$, and $s$ quarks to be massless. This leads to a simplification due to the unitarity of the CKM matrix. Unitarity implies that
\begin{align}
\label{ckm}
V^*_{ts}\,V_{td} + V^*_{cs}V_{cd}+V^*_{us}V_{ud}=0\,.
\end{align}
Due to its small size, $V^*_{ts}\,V_{td}$ can be neglected here, and thus using Eq.~(\ref{ckm}) on Eq.~(\ref{heff}) and expanding on $x_c$ leads to 
\begin{align}
\label{cancelXu}
V^*_{cs}V_{cd}\,X (x_c)+ V^*_{us}V_{ud}\,X (x_u)\simeq V^*_{cs}V_{cd}\big( X (x_c) - X(0) \big) = V^*_{cs}V_{cd}\big( X' (0)\cdot x_c + \dots\big)\,,
\end{align}
where higher powers in $x_c$ are neglected. This means that in the diagrams of Figure~\ref{boxandpengdiags}, and in their QCD corrections, one need only keep terms proportional to $m_c^2$. Perturbative QCD effects lead to large logarithms of the form $L=\ln(\mu_c^2/\mu_W^2)$, which need to be resummed using Renormalization Group (RG) equations and an effective theory framework. This calculation was performed at $\mathcal{O}(\alpha_s^n L^{n+1})$ (LO) in Refs.~\cite{Vainshtein:1976eu,Ellis:1982ve,Dib:1989cc}. A study at $\mathcal{O}(\alpha_s^n L^{n})$ (NLO) was performed in Refs.~\cite{Buchalla:1993wq,Buchalla:1998ba}, which was extended to $\mathcal{O}(\alpha_s^n L^{n-1})$ (NNLO) in Ref.~\cite{Buras:2006gb}. 
The NLO calculation did not take into account the contributions from the anomalous penguin diagrams, such as the ones shown in Figure~\ref{anomdiags}. These diagrams were considered in the NNLO calculation, but were not computed properly.
A Chern-Simons operator was introduced, which supposedly originated from integrating out the top quark in the diagrams in Figure~\ref{toptrdiags}.
However, these diagrams actually vanish in the large $m_t$ limit\cite{Collins:1978wz,Chetyrkin:1993hk}, so the Wilson coefficient for the Chern-Simons operator vanishes as well, and this operator would make no contribution.\footnote{As this work was being finished, we learned that the authors of Ref.~\cite{Buras:2006gb} had prepared an Erratum to their original paper in which they recalculated the contribution of the anomalous penguin diagrams. We discuss the Erratum in section~\ref{comp}}

The goal of this paper is to reevaluate the penguin charm contributions to $K^+\to \pi^+\nu\bar{\nu}$,  checking the results presented in Ref.~\cite{Buras:2006gb}.
In section~\ref{secdefs} we briefly present our renormalization scheme, determined by our definition of the axial-vector current and off-diagonal renormalization. In section~\ref{secpert} we present our purely perturbative results for the anomalous diagrams. In section~\ref{secRGNA} we perform the RG treatment of non-anomalous diagrams, which pervades that of the anomalous ones. Then in sections \ref{secRGtop} and \ref{secRGbottom} we treat the anomalous diagrams involving a top and a bottom triangle, respectively. Finally, in section~\ref{numres} we present and discuss our numerical results and conclude in section~\ref{conclusion}.

\section{Definition of the axial-vector current and off-diagonal renormalization}
\label{secdefs}

We perform our calculations using dimensional regularization and the $\overline{MS}$ renormalization scheme. Beyond this, our scheme is determined by our choice of renormalization for the axial-vector current and off-diagonal corrections to the quark propagator.

\subsection{The axial-vector current}
\label{defLarin}

In dimensional regularization special attention must be paid to the definition of $\gamma_5$. Whenever it appears in an open fermion line we can safely use its naive definition, but in closed loops we use the original definition by 't Hooft and Veltman \cite{'tHooft:1972fi},
\begin{equation}
\label{gamma5}
\gamma_5=i\frac{1}{4!}\epsilon_{\nu_1\nu_2\nu_3\nu_4}\gamma_{\nu_1}\gamma_{\nu_2}\gamma_{\nu_3}\gamma_{\nu_4}\,.
\end{equation}
The Levi-Civita $\epsilon$ tensor is unavoidably a four-dimensional object and thus is taken outside the R-operation where a D-dimensional object can be safely considered as a four-dimensional one. The gamma matrices are taken as D-dimensional inside the R-operation. With this definition $\gamma_5$ no longer anticommutes with D-dimensional $\gamma_{\mu}$. For this reason we use the definition of the axial-vector current presented in Ref.~\cite {Larin:1993tq},
\begin{equation}
\label{defAi}
A^{q,0}=\frac{1}{2}\overline{\psi}_q\left( \gamma_{\mu}\gamma_5 - \gamma_5\gamma_{\mu}\right) \psi_q\to i\frac{1}{3!}\epsilon_{\mu\nu_1\nu_2\nu_3}\overline{\psi}_q\gamma_{\nu_1}\gamma_{\nu_2}\gamma_{\nu_3}\psi_q\,.
\end{equation}
The renormalized current is defined as $A_R^q=\xi_I^{\rm NS}\left( Z^{\rm NS}\,A_0^q + Z^{\psi}\,A_0^{\rm S}\right)$, where $A^{\rm S}=\sum_{q}A^q$\cite{Larin:1993tq}. The renormalization constant $Z^{\rm NS}$ cancels divergences arising from non-anomalous diagrams, whereas $Z^{\psi}$ deals with anomalous ones, and thus begins at order $\alpha_s^2$. The coefficient $\xi_I^{\rm NS}$ is a finite piece that is determined by requiring that the matrix elements of the renormalized non-singlet axial-vector and vector vertices coincide,
\begin{equation}
\xi_I^{\rm NS} \langle \overline{\psi} \gamma_{\mu}\gamma_5T^a\psi\rangle^{R}=\langle \overline{\psi}\gamma_{\mu}T^a\psi\rangle^{R}\gamma_5\,,
\end{equation}
where $T^a$ is the generator of a flavor group.
This effectively restores the anticommutativity of the $\gamma_5$ matrix for the non-singlet vertex, and so the standard Ward identities as well. This also means that the non-singlet axial-vector current defined in this way has zero anomalous dimension.

\subsection{Off-diagonal renormalization}

\begin{figure}
\centering
\includegraphics[width=0.25\columnwidth]{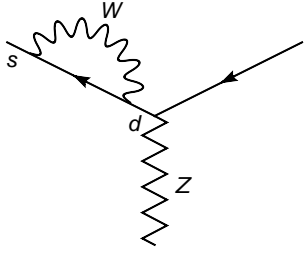}
\caption{\label{pictred}
An example of a reducible penguin-type diagram.}
\end{figure}

The $W$ and $\phi$ bosons can lead to flavor-changing corrections to the quark propagator, which in turn imply the appearance of reducible contributions to the penguin-type diagrams, such as the one in Figure~\ref{pictred}.
In order to avoid these diagrams, we choose to renormalize the left-handed doublets in the following way \cite{Voloshin:1976in,Buchalla:1992zm,Misiak:2004ew},
\begin{equation}
\left(
\begin{array}{l}
u\\
d'\\
\end{array}
\right)^0_L
= Z_1^{1/2}
\left(
\begin{array}{l}
u\\
d'\\
\end{array}
\right)^R_L
\,,
\quad
\left(
\begin{array}{l}
c\\
s'\\
\end{array}
\right)^0_L
= Z_2^{1/2}
\left(
\begin{array}{l}
c\\
s'\\
\end{array}
\right)^R_L
\,,
\end{equation}
where $d' = d \cos \theta_W + s \sin \theta_W$ and $s' = -d \sin \theta_W + s \cos \theta_W$. This leads to the following off-diagonal counterterm in the lagrangian,
\begin{eqnarray}
\label{Lct}
\mathcal{L}_{ct} &=& \left[\overline{s}_L\left( i \slashed{\partial} + g_s t^a G_{\mu}^a\gamma^{\mu} +  \frac{g_{EW}}{\cos\theta_W}\gamma^{\mu}\left(I_3^d -\sin^2\theta_W\, e_d\right)Z_{\mu}\right)d_L + (s\leftrightarrow d)\right]\nonumber\\
&&\times(Z_1-Z_2)\sin\theta_W\cos\theta_W\,,
\end{eqnarray}
which contains the couplings to a gluon $G^a_{\mu}$, with the strong coupling constant $g_s$ and the $SU(3)$ generator $t^a$, and the coupling to a $Z$ boson, with the electroweak coupling constant $g_{EW}$. $I_3^q=(+1/2,-1/2)$ and $e_q=(+2/3,-1/3)$ are the electroweak isospin and electric charge of up- and down-type quarks, respectively.

We set the value of $(Z_1-Z_2)$ by requiring the cancellation of off-diagonal self-energy corrections,
\begin{equation}
\parbox{40mm}{\includegraphics[width=0.22\columnwidth]{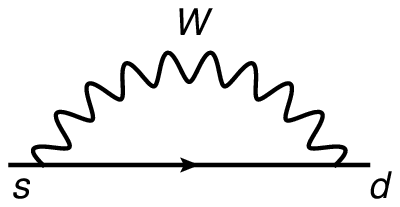}}\quad + \quad \parbox{40mm}{\includegraphics[width=0.22\columnwidth]{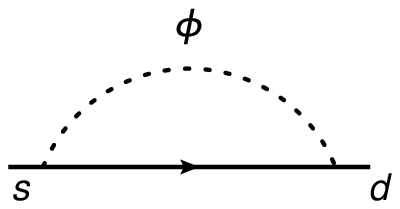}}\quad +  \quad\parbox{40mm}{\includegraphics[width=0.22\columnwidth]{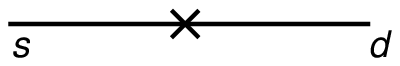}}\quad = 0\,,
\end{equation}
plus corrections in $\alpha_s$. This fixes also the off-diagonal gluon and Z-vertex counterterms.

\begin{figure}[h]
\centering
\subfloat{
 \includegraphics[width=.19\linewidth]{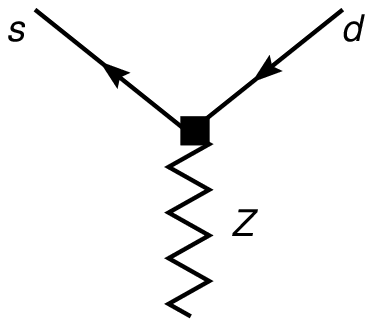}
 }
 \subfloat{
 \includegraphics[width=.19\linewidth]{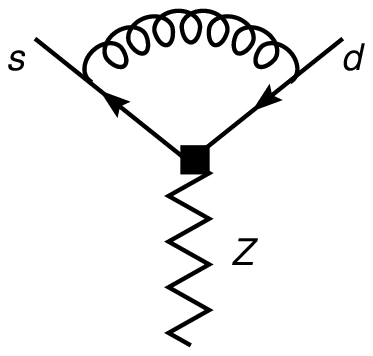}
}
\caption{Off-diagonal Z-vertex counterterms at zero and one loops.\label{cntZvrtx}}
\end{figure}

\begin{figure}[h]
\centering
\subfloat{
 \includegraphics[width=.19\linewidth]{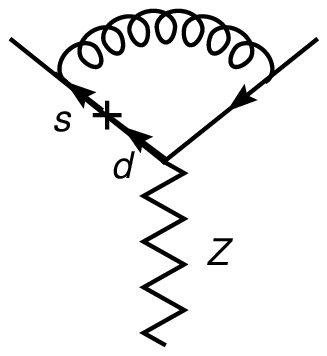}
 }
 \subfloat{
 \includegraphics[width=.19\linewidth]{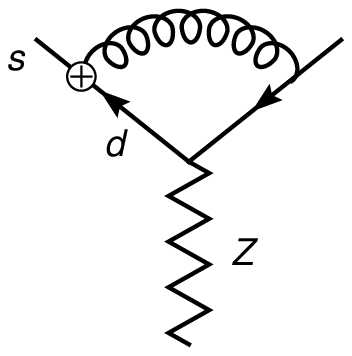}
}
\caption{Two irreducible diagrams involving off-diagonal counterterms.\label{cntpgvrtx}}
\end{figure}

\begin{figure}[hhh!]
\centering
\includegraphics[width=0.70\columnwidth]{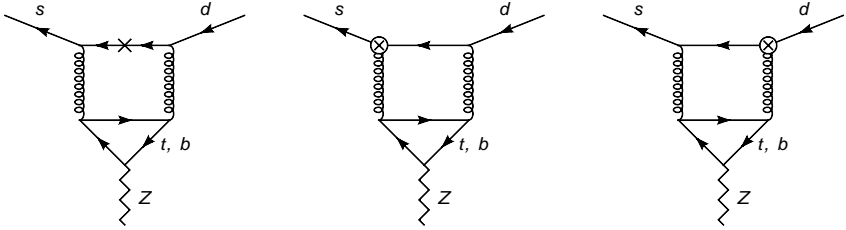}
\caption{\label{pictcount}
The counterterm diagrams for the anomalous case.}
\end{figure}

In the case of non-anomalous penguin diagrams, only the Z-vertex counterterms, like the ones shown in Figure~\ref{cntZvrtx}, survive. By definition, reducible diagrams like the one in Figure~\ref{pictred} are cancelled by the insertion of the off-diagonal propagator counterterm in the external lines.
Then we can have irreducible diagrams like the two shown in Figure~\ref{cntpgvrtx}. However, because the factor $i\slashed{\partial}$ in the propagator counterterm kills the denominator of one fermion propagator (and the factors of $i$ introduce a sign), the sum of these two diagrams is zero. In the non-anomalous case we always have the same number of diagrams involving off-diagonal propagator or gluon counterterms, so they all cancel out against each other. The irreducible counterterm diagrams in the anomalous case are shown in Figure~\ref{pictcount}. Here the balance is broken, and one diagram with an off-diagonal gluon counterterm survives.

\section{Perturbative results for anomalous diagrams}
\label{secpert}

Since the anomalous diagrams where initially computed incorrectly in Ref.~\cite{Buras:2006gb}, we
will show their calculation in detail, and begin by the perturbative results from their diagrammatic
calculations. We used the program \texttt{qgraf}~\cite{Nogueira:1991ex} to generate all
of the diagrams, and the packages \texttt{q2e} and
\texttt{exp}~\cite{Harlander:1997zb,Seidensticker:1999bb} to express them as a series of
vertices and propagators that can be
read and evaluated by the \texttt{FORM}~\cite{Vermaseren:2000nd} package
\texttt{MATAD~3}~\cite{Steinhauser:2000ry}.

For greater convenience when resumming logarithms later on, we will express the results
in terms of the operator $Q_{\nu}(\mu)$, defined as
\begin{align}
Q_{\nu}(\mu)=\frac{m_c^2(\mu)}{g_s^2(\mu)\mu^{2\epsilon}}\,\bar{s}\gamma^{\mu}(1 -
\gamma_5)d\,, \label{Qnu}
\end{align}
and write the expression for the charm contribution to $\mathcal{H}_{\rm eff}$ as
\begin{align}
\mathcal{H}^c_{\rm eff}=&\frac{G_F}{\sqrt{2}}\frac{\alpha}{2\pi\sin^2\theta_W}\frac{\pi^2}{2M_W^2}V^*_{cs}V_{cd}\,\mathcal{H}_\nu\otimes\sum_{l=e,\mu,\tau} \bar{\nu}_l\gamma_{\mu}(1 -
\gamma_5)\nu_l
\nonumber\\
=&\frac{g_{EW}^4}{128\,M_W^4}V^*_{cs}V_{cd}\,\mathcal{H}_\nu\otimes \sum_{l=e,\mu,\tau}
\bar{\nu}_l\gamma_{\mu}(1 - \gamma_5)\nu_l\,,\label{heff2}
\end{align}
where we have defined $\mathcal{H}_\nu= D\,Q_\nu(\mu)$ and split up the global factor comming from the external neutrinos.
We factored out the factor $\pi^2/(2M_W^2)$ so that the coefficient $D$ will have the same normalization as
the decoupling coefficients in Refs.~\cite{Buchalla:1993wq,Buras:2006gb}. The relation between X, defined in Eq.~(\ref{heff}), and D is simply $X = \pi^2/(2M_W^2)\,m_c^2/g_s^2\,D$.

Because of the different logarithms that appear in them, we divide the coefficient $D$ in
different contributions, $D=D_{W,t}+D_{W,b}+D_\phi$. Here $D_{W,t}$ stands for diagrams
with a $W$ boson and a top triangle, $D_{W,b}$ for diagrams with $W$ and a bottom
triangle, and $D_\phi$ contains all anomalous diagrams with Goldstone bosons
( $\ell_{\mu/m_X}=\ln[\mu^2/m_X^2(\mu)]$ )
\begin{align}
D_{W,t}=&~ \Big(a^{(6)}(\mu)\Big)^3\,12\,C_F\Big[\big( 1 -4\ell_{\mu/m_t}\big)\big(1 +
\ell_{\mu/M_W} - \ell_{\mu/m_c}\big)\Big] ~,\label{pert1}
\\ D_{W,b}=&~ \Big(a^{(6)}(\mu)\Big)^3 4\,C_F\Big[ 6\ell_{\mu/M_W}^2 +
6\ell_{\mu/m_b}^2 -12 \ell_{\mu/m_b}\ell_{\mu/m_c}           +9\ell_{\mu/M_W} +3\ell_{\mu/m_c} +36+
2\pi^2\Big] ~, \label{pert2}
\\
D_{\phi}=&~ -\Big(a^{(6)}(\mu)\Big)^3\,C_F\Big[ 6\ell_{\mu/m_t}^2 + 6
\ell_{\mu/M_W}^2           - 12 \ell_{\mu/m_t}( 3+\ell_{\mu/M_W})+ 36 \ell_{\mu/M_W} + 75 +
2\pi^2\Big] ~,\label{pert3}
\\
D=&\Big(a^{(6)}(\mu)\Big)^3 3\, C_F\Big[ 8 \ell_{M_W/m_b}^2 - 2 \ell_{M_W/m_t}^2- 16 \ell_{M_W/m_b} \ell_{M_W/m_c}  -
 4 \ell_{M_W/m_t} (1 - 4 \ell_{M_W/m_c}) 
\nonumber\\
&\hspace{2.8cm}+ 2\,\pi^2+ 27 \Big]~,\label{pertTotal}
\end{align} 
with $a^{(nf)}=\alpha_s^{(n_f)}/(4\pi)$ being the QCD coupling constant in the
effective $n_f$ flavor theory and $C_F=4/3$. One can see by inspecting Eqs.~(\ref{pert1})-(\ref{pert3}) that the explicit $\ell_\mu$ dependence cancels separately in $D^{\phi}$ and the sum $D^{W,t}+D^{W,b}$.

As they are, the results from the diagrams involving the
$W$ boson are of little use because of the large logarithms present in them, and an RG
treatment is called for.
However, as we can see in (\ref{pert3}) the Goldstone contribution only contains the $W$
and top quark scale, which are of similar size. We won't treat this
contribution when we proceed with the RG improvement in the effective field theory
language.


\section{RG improvement: non-anomalous diagrams}
\label{secRGNA}

\begin{figure}
\centering
\parbox{40mm}{\includegraphics[width=0.25\columnwidth]{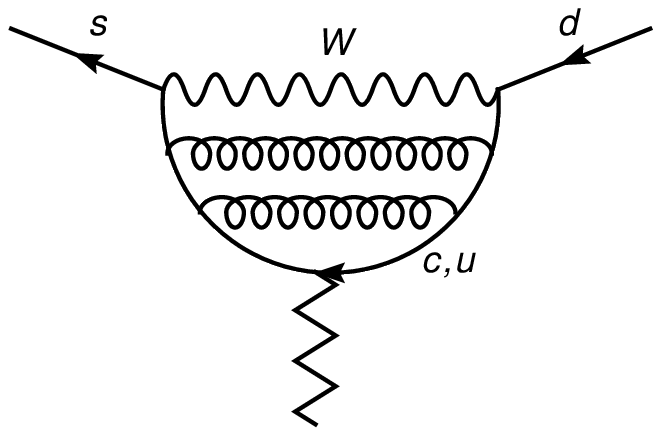}}\qquad\quad \parbox{40mm}{\includegraphics[width=0.25\columnwidth]{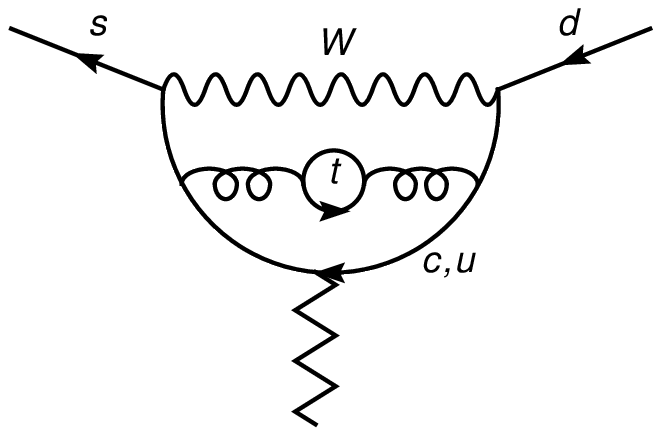}}
\caption{\label{NANNLO}
Two non-anomalous penguin diagrams at NNLO.}
\end{figure}

Part of the goal of this paper is to check the results for non-anomalous penguin diagrams from Ref.~\cite{Buras:2006gb}.
These diagrams are always present as subgraphs in the anomalous case. In particular, we will need the Wilson coefficients and anomalous dimensions related to the decoupling of the $M_W$ scale up to NNLO, that is up to diagrams like the ones in Figure~\ref{NANNLO}.
It is useful then to present first the full RG improvement of the non-anomalous case at NNLO, and introduce the elements that we will need throughout.

All penguin diagrams originate from the $Z$ current, where the $Z$ coupling to a fermion $q$ is defined as
\begin{align}
Q_Z^q=v_q\,V^q+a_q\,A^q=2\,\Big( I_3^q  - 2\,e_q\,\sin^2\theta_W\Big)\,V^q - 2\,I_3^q \,A^q\,,
\end{align}
with the vector and axial-vector current
\begin{align}
V^q=\overline{\psi}_q\,\gamma^{\mu}\,\psi_q \qquad\quad{\rm and}\qquad\quad A^q=\overline{\psi}_q\,\gamma^{\mu}\gamma_5\,\psi_q~.
\end{align}
Since for the vector current the anomalous diagrams vanish and we want to use some results of the non-anomalous calculation later on, we look at the vector and axial-vector current separately. In the non-anomalous diagrams only the $Z$ coupling to the four lightest quarks ($d$, $u$, $s$ and $c$) is present, so it is convenient to define the non-anomalous current as
\begin{align}
 A^{\rm NA}=-A^c-A^u+A^s+A^d~,
\end{align}
which is the sum of all axial-vector parts of the $Z$ current, the signs being determined by the prefactor $-2\,I_3^q$. In our renormalization scheme, the operators $A^s$ and $A^d$ only enter in anomalous diagrams, to cancel the ones coming from $A^c$ and $A^u$.
Since the axial-vector current appears only on open fermion lines in non-anomalous diagrams, we could (and did) use the naive definition of $\gamma_5$ in their computation. Using the definition in Eq.~(\ref{gamma5}) leads in principle to different results: the difference between inserting one $\gamma_5$ or the other on a fermion line is of $\mathcal{O}(\epsilon)$. If this line is part of a loop diagram with a pole $1/\epsilon$, as in the penguin diagrams involving a Goldstone boson, the difference will turn finite. However, since this difference stems from a residue, it will be local, and thus it can be cancelled by adding a finite counterterm proportional to $Q_{\nu}$. If one does this consistently at every order the results with a naive $\gamma_5$ or with 't Hooft and Veltman's $\gamma_5$ can be made to agree.

We will now present the decoupling of the three heavy scales in our problem, $m_t$, $M_W$, and $m_b$. Each decoupling will be assumed to take place at the appropriate scale $\mu$ close to the heavy scale in particular, and we will present the RG evolution that can make this possible afterwards.

\subsection{Decoupling at the top scale}

The first scale to separate is that of the top quark, which in this case only enters through gluon self-energy corrections, as in Figure~\ref{NANNLO}. Since we neglect all power corrections involving the top mass, this decoupling is trivial, as it is given solely by the decoupling relations from $n_f=6$ to $n_f=5$ in all the elements involved in this calculation, namely $a$, the masses, the gauge parameter $\xi$, and the wave functions. These relations can be found in \cite{Chetyrkin:1997un}.

\subsection{Decoupling at the $W$ scale}

\begin{figure}[t]
\centering
\subfloat[]{
\includegraphics[width=.25\linewidth]{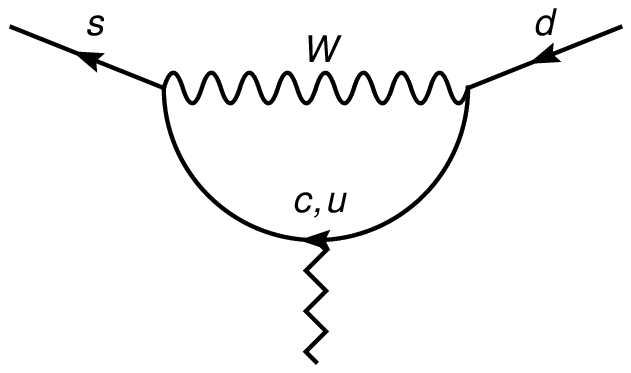}
\label{LOPeng}
}
\hspace{1cm}
\subfloat[]{
 \includegraphics[width=.2\linewidth]{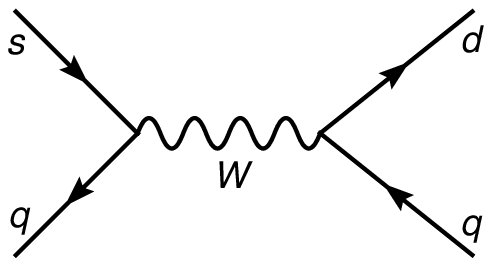}
\label{4qLO}
}
\caption{Leading order examples of a penguin-diagram (a) and a four-quark subdiagram (b).}
\end{figure}

Next one needs to consider the decoupling of the $W$ boson. Let us therefore look at the leading order penguin-diagram depicted in Figure \ref{LOPeng}. When considering an expansion of the diagram in terms of $m_c^2/M_W^2$ there are two distinct cases. The first one is when the loop momentum going through the $u(c)$ line is of the order of $M_W$. Then the whole diagram shrinks to a point and can be represented by the local operator $Q_\nu$ defined in Eq.~(\ref{Qnu}). The second case is when the loop momentum going through the $u(c)$ line at the $Z$ vertex is of the order of $m_c$, and $M_W$-sized loop momenta run only through a subdiagram containing the $W$ line, like the one shown in Figure~\ref{4qLO}. The decoupling of the $W$ in the subdiagram can be described by the effective Hamiltonian
\begin{align}
\label{Leff}
 \mathcal{H}_{\rm eff}^{W,q} =\frac{g_{EW}^2}{8M_W^2}\left( C_+\,Q^q_+ + C_-\,Q^q_-\right)~,
\end{align}
with the definition of the effective operators ($\alpha$ and $\beta$ are color indices)
\begin{align}
Q_{\pm}^q=\frac{1}{2}\Big[ \big(\bar{s}^{\alpha}\gamma^{\mu}(1-\gamma_5)q^{\alpha}\big)\big(\bar{q}^{\beta}\gamma_{\mu}(1-\gamma_5)d^{\beta}\big) \pm \big(\bar{s}^{\alpha}\gamma^{\mu}(1-\gamma_5)q^{\beta}\big)\big(\bar{q}^{\beta}\gamma_{\mu}(1-\gamma_5)d^{\alpha}\big)\Big]~.\label{defQpm}
\end{align}
The combination of color indices in the definition of $Q_{\pm}^q$ is chosen so that these operators will not mix under renormalization.
The value of the coefficients will be given later on. After decoupling $W$ in this subdiagram, the whole diagram is then represented by the bilocal operators
\begin{align}
 V_{\pm}&=-i\,\int \mathrm{d}^4x~\mathcal{T}\Big\{ \sum_{q=d,u,s,c} \big(Q^c_{\pm}(x)-Q^u_{\pm}(x)\big)\,v_q\,V^q(0)\Big\}~,
 \\
 A^{\rm NA}_{\pm}&=-i\,\int \mathrm{d}^4x~\mathcal{T}\Big\{ \big(Q^c_{\pm}(x)-Q^u_{\pm}(x)\big)
                      \,A^{\rm NA}(0)\Big\}~,
\end{align}
for the vector and axial-vector current, respectively. The minus sign between $Q^c_{\pm}$ and $Q^u_{\pm}$ comes from the GIM mechanism discussed in Eq.~(\ref{cancelXu}).

Besides the operators $Q_{\pm}^q$ one must also consider a set of evanescent operators $Q_{E_i}^q$ (and the corresponding bilocal operators). We take the definitions from Ref.~\cite{Buras:2006gb} (modified by a normalization factor), which we present in Appendix \ref{appe}. Evanescent operators vanish at four dimensions, but yield non-zero contributions when inserted in loop diagrams with poles in $\epsilon$. We choose a renormalization scheme for them in which their matrix elements vanish and they do not mix with physical operators \cite{Dugan:1990df}. The physical operators still mix with them, though, so the evanescents contribute to the anomalous dimensions of $Q_{\pm}^q$, and, inserted into bilocal operators, to those of $V_{\pm}$ and $A^{\rm NA}_{\pm}$.

Putting everything together, we describe the non-anomalous penguin-diagrams with a vector and axial-vector current with the Hamiltonian 
\begin{align}
 \mathcal{H}^ {\rm{NA}}_{\nu, \rm eff} =&~ C_+\left[V_+ + A^{\rm NA}_+\right] + C_-\left[V_-  + A^{\rm NA}_-\right]+ C^{\rm NA}_\nu\,Q_\nu~,\label{defHNA}
 \end{align}
Here we have omitted the normalization and neutrino factors, which are the same as the ones multiplying $H_{\nu}$ in Eq.~(\ref{heff2}).
The coefficients in $\mathcal{H}_{\nu,\rm{eff}}^{\rm{NA}}$ are ( $\ell_{\mu_W/M_W}=\ln[\mu_W^2/M_W^2]$ )
\begin{align}
 C_{+}(\mu_W)=&~1+a^{(5)}(\mu_W)\Big[\tfrac{11}{3}+2\,\ell_{\mu_W/M_W} \Big]
 \nonumber\\
 &~+\Big(a^{(5)}(\mu_W)\Big)^2\Big[ \ell_{\mu_W/M_W}^2\big( 13-\tfrac{2}{3}n_f \big) +\tfrac{5}{18}\ell_{\mu_W/M_W}(159-8\,n_f)
 \nonumber\\
 &~\qquad-\tfrac{1}{18}(55+24\,\zeta_2)\,n_f+ 26\,\zeta_2+\tfrac{21709}{1800}\Big]~,
 \\
 C_{-}(\mu_W)=&~1-a^{(5)}(\mu_W)\Big[\tfrac{22}{3}+4\,\ell_{\mu_W/M_W} \Big]
 \nonumber\\
 &~+\Big(a^{(5)}(\mu_W)\Big)^2\Big[ \ell_{\mu_W/M_W}^2\big( -14+\tfrac{4}{3}n_f \big)\ -\tfrac{5}{9}\ell_{\mu_W/M_W}(105-8\,n_f) 
 \nonumber\\
 &\qquad+\tfrac{1}{9}(55+24\,\zeta_2)\,n_f - 28\,\zeta_2-\tfrac{92443}{900}\Big]~,
\end{align}
\begin{align}
C^{\rm NA}_{\nu}(\mu_W)=&~a^{(5)}(\mu_W)\,8 \Big[ \ell_{\mu_W/M_W}+2\Big]+ \Big(a^{(5)}(\mu_W)\Big)^2\,\tfrac{16}{3} 
\Big[12\,\ell_{\mu_W/M_W}^2+ 34\,\ell_{\mu_W/M_W}  + 24\,\zeta_2+33\Big] \nonumber\\
&+\Big(a^{(5)}(\mu_W)\Big)^3\,\Big[  - \tfrac{64}{3} \ell_{\mu_W/M_W}^3 (-24 + n_f) + \ell_{\mu_W/M_W}^2 \left(3408 - \tfrac{1024}{9}n_f\right)
\nonumber\\
&\hspace{2.8cm} - \tfrac{16}{9} \ell_{\mu_W/M_W} \Big(-5270 - 1728\,\zeta_2 + n_f (161 + 72\,\zeta_2) + 468\,\zeta_3\Big)
\nonumber\\
 &\hspace{2.8cm} - 416\,\zeta_4 - \tfrac{8896}{3}\zeta_3 + 6816\,\zeta_2 - \tfrac{64}{9} n_f (49 + 32\,\zeta_2 - 12\,\zeta_3)  \nonumber\\
    &\hspace{2.8cm}
    + \tfrac{7995148}{675}\Big]~.
\end{align}
The bilocal operators $V_\pm$ do not mix under the RG evolution with other operators and their matrix elements vanish. Therefore, we don't have to consider the vector current any longer, its only effect being inside the decoupling constant $C^{\rm NA}_\nu$.

\subsection{Decoupling at the bottom scale}

First note that when decoupling a heavy quark, the non-singlet vector or axial-vector current is the same in the full and effective theory due to Ward identities\cite{Collins:1978wz,Chetyrkin:1993hk,Chetyrkin:1993jm}. As a consequence, the non-anomalous current is the same in the five and four flavor theory ($[A^{\rm NA}]^{(5)}=[A^{\rm NA}]^{(4)}$). Also, the decoupling coefficient for the $Q_\nu$ operator stems purely from the prefactor $m_c^2/g_s^2$,
\begin{align}
 \Big[Q_{\nu}\Big]^{(5)}&= B_\nu\,\Big[Q_{\nu}\Big]^{(4)}~, \label{Qnubdec}
\end{align}
with ( $\ell_{\mu_b/m_b}=\ln[\mu_b^2/m_b^2(\mu_b)]$  )
\begin{align}
 B_{\nu}(\mu_b)=& 1-\frac{2}{3}a^{(5)}(\mu_b)\,\ell_{\mu_b/m_b} - \frac{2}{27}\,\Big(a^{(5)}(\mu_b)\Big)^2\Big[ 30\,\ell_{\mu_b/m_b}^2 + 39\,\ell_{\mu_b/m_b} + 56 \Big]~.
\end{align}
The decoupling equations for the local operators $Q_{\pm}^q$ are
\begin{align}
 \Big[Q^q_{\pm}\Big]^{(5)}&= B_\pm\,\Big[Q^q_{\pm}\Big]^{(4)}~, \label{Qpmbdec}
\end{align}
with
\begin{align}
 B_{+}(\mu_b)=&1-\frac{1}{54}\,\Big(a^{(5)}(\mu_b)\Big)^2\Big[ 36\,\ell_{\mu_b/m_b}^2 +12\,\ell_{\mu_b/m_b}+59\Big]~,
 \\
 B_{-} (\mu_b) =&1+\frac{1}{27}\,\Big(a^{(5)}(\mu_b)\Big)^2 \Big[ 36\,\ell_{\mu_b/m_b}^2 +12\,\ell_{\mu_b/m_b}+59\Big]~.
\end{align}
These results then can be used to get the decoupling coefficients for the bilocal operators, whose decoupling equations read
\begin{align}
 \Big[A^{\rm NA}_{\pm}\Big]^{(5)}&= B_\pm\,\Big[A^{\rm NA}_{\pm}\Big]^{(4)} + B^{\rm NA}_{\pm,\nu}\,\Big[Q_{\nu}\Big]^{(4)}~,
\label{ANApmbdec}
\end{align}
with
\begin{align}
 B^{\rm NA}_{+,\nu} (\mu_b) =&\frac{8}{9}\,\Big(a^{(5)}(\mu_b)\Big)^3 \Big[ 12\,\ell_{\mu_b/m_b}^3-96\,\ell_{\mu_b/m_b}^2 + 261\,\ell_{\mu_b/m_b}+96\,\zeta_3-\tfrac{1279}{3}\Big]~,
 \\
 B^{\rm NA}_{-,\nu,u} (\mu_b) =&-\frac{8}{9}\,\Big(a^{(5)}(\mu_b)\Big)^3 \Big[ 12\,\ell_{\mu_b/m_b}^3-96\,\ell_{\mu_b/m_b}^2 + 241\,\ell_{\mu_b/m_b}+96\,\zeta_3-\tfrac{1231}{3}\Big]~.
\end{align}

\subsection{Decoupling at the charm scale}

At the charm scale one matches finally the operator $A^{\rm NA}_{\pm}$ to $Q_\nu$ and finds the coefficient $X(x_c)$, cf. Eq.~(\ref{heff}). 
The operator multiplying $X$ has no anomalous dimension, which means that $X$ is independent of $\mu$. In this case it is unnecessary to move to a three-flavor theory, and one stays in the four-flavor one,
\begin{align}
\Big[A^{\rm NA}_{\pm}\Big]^{(4)}=C^{\rm NA}_{\pm,\nu}\,\Big[Q_{\nu}\Big]^{(4)}~, \label{ANAcmatching}
\end{align}
with the coefficients ( $\ell_{\mu_c/m_c}=\ln[\mu_c^2/m_c^2(\mu_c)]$  )
\begin{align}
 C^{\rm NA}_{+,\nu}(\mu_c)=&16\,a^{(4)}(\mu_c) \Big[1 - \ell_{\mu_c/m_c}\Big] - \Big(a^{(4)}(\mu_c)\Big)^2 \Big[48 \ell_{\mu_c/m_c}^2 - 80 \ell_{\mu_c/m_c} - 44 \Big]~,\label{CNA+}\\
C^{\rm NA}_{-,\nu}(\mu_c)=&-8\,a^{(4)}(\mu_c) \Big[1 - \ell_{\mu_c/m_c}\Big] + \frac{4}{3}\,\Big(a^{(4)}(\mu_c)\Big)^2 \Big[ 36 \ell_{\mu_c/m_c}^2+ 36 \ell_{\mu_c/m_c}-21\Big]~.\label{CNA-}
\end{align}
After showing the decouplings at different scales, we now show how the operators can be evolved between these scales.

\subsection{Running to different scales}

The evolution of the operator $Q_{\nu}$ is given by
\begin{align}
\label{evnu}
\mu^2\frac{d}{d\mu^2}Q_{\nu}(\mu)=\gamma_{\nu}\, Q_{\nu}(\mu)~,
\end{align}
where $\gamma_{\nu}$ is easily determined from the renormalization constants of $m_c$ and $\alpha_s$. Defining $d\mu^2/\mu^2 a=a\beta(a)$,
\begin{eqnarray}
\gamma_{\nu}=2(\gamma_m-\beta)&=&a \Big[3 - \frac{2}{3}n_f\Big] - \frac{2}{9} a^2 \Big[147 + 37\,n_f\Big]\nonumber\\
&& + 
 \frac{1}{162} a^3 \Big[-173259 - 18705\,n_f + 1535\,n_f^2 + 17280\,\zeta_3 n_f \Big]\,.
\end{eqnarray} 
The solution of Eq.~(\ref{evnu}) is
\begin{align}
 Q_{\nu}(\mu)
 =U_\nu(\mu,\mu_0)\,Q_{\nu}(\mu_0)
 =\rm{exp}\bigg(\int_{a(\mu_0)}^{a(\mu)}\frac{dz}{z}\frac{\gamma_{\nu}(z)}{\beta(z)}\bigg)\,Q_{\nu}(\mu_0)\,.\label{Unu}
\end{align}

Considering the RG equations for the local operators $Q_\pm$ and $A^{\rm{NA}}$
\begin{align}
 \mu^2\frac{d}{d\mu^2}Q^q_{\pm}(\mu)=&~\gamma_\pm\,Q^q_{\pm}(\mu)~,
 \\
 \mu^2\frac{d}{d\mu^2}A^{\rm{NA}}(\mu)=&~0~,
\end{align}
one gets for the bilocal operators
\begin{align}
\label{evQP}
\mu^2\frac{d}{d\mu^2} A^{\rm NA}_{\pm}(\mu)=\gamma_{\pm}\, A^{\rm NA}_{\pm}(\mu) + \gamma_{\pm,\nu}^{\rm NA}\, Q_{\nu}(\mu)~.
\end{align}
The anomalous dimensions are
\begin{align}
 \gamma_+=&~ -2\,a + a^2\,\Big[ \tfrac{7}{2}-\tfrac{2}{9}n_f \Big] + a^3\,\Big[ -\tfrac{275267}{300}+\tfrac{52891}{1350}\,n_f
       +\tfrac{130}{81}\,n_f^2+\big(336+\tfrac{80}{3}\,n_f\big)\zeta_3 \Big]~,
\\
 \gamma_-=&~ 4\,a + a^2\,\Big[ 7+\tfrac{4}{9}n_f \Big] + a^3\,\Big[ -\tfrac{12297}{50}+\tfrac{31343}{675}\,n_f-\tfrac{260}{81}\,n_f^2-\big(336+\tfrac{160}{3}\,n_f\big)\zeta_3 \Big]~,
\end{align}
and
\begin{align}
 \gamma_{+,\nu}^{\rm NA}=&~ -16\,a - 144\,a^2 + a^3\,\Big[ -\tfrac{1060082}{225}+144\,n_f+896\,\zeta_3 \Big]~,
\label{anomdimNA+}
\\
 \gamma_{-,\nu}^{\rm NA}=&~ 8\,a + 208\,a^2 + a^3\,\Big[ \tfrac{879586}{225}-144\,n_f-64\,\zeta_3 \Big]~.
\label{anomdimNA-}
\end{align}
The factor $1/g_s^2$ was introduced in the definition of $Q_{\nu}$, cf. Eq.~(\ref{Qnu}), so that the anomalous dimensions $\gamma^{\rm NA}_{\pm,\nu}$ would begin at order $a$, and not $a^0$.
The solution of Eq.~(\ref{evQP}) is
\begin{align}
 \label{solevQP}
 A^{\rm NA}_{\pm}(\mu)= U_{\pm}(\mu,\mu_0)\,A^{\rm NA}_{\pm}(\mu_0) + U_{\pm,\nu}^{\rm NA}(\mu,\mu_0)\,Q_\nu(\mu_0)~,
\end{align}
where
\begin{align}
 U_{\pm}(\mu,\mu_0)=&~\rm{exp}\bigg(\int_{a(\mu_0)}^{a(\mu)}\frac{dz}{z}\frac{\gamma_{\pm}(z)}{\beta(z)}\bigg)~, \label{Upm}
 \\
 U_{\pm,\nu}^{\rm NA}(\mu,\mu_0) =&~ \rm{exp}\bigg(\int_{a(\mu_0)}^{a(\mu)}\frac{dz}{z}\frac{\gamma_{\pm}(z)}{\beta(z)}\bigg) \cdot
  \int_{a(\mu_0)}^{a(\mu)}\frac{dz}{z}\frac{\gamma_{\pm,\nu}^{\rm NA}(z)}{\beta(z)}
 \nonumber \\
  &\hspace{2cm}\cdot\rm{exp}\bigg(\int_{a(\mu_0)}^{z}\frac{dz'}{z'}\frac{\gamma_{\nu}(z')-\gamma_{\pm}(z')}{\beta(z')}\bigg)~. \label{Upmnu}
\end{align}

\subsection{Result}
\label{NAres}
Using all the decoupling coefficients and evolution equations presented in the previous sections one can find the final result for the non-anomalous diagrams. 
We start by Eq.~(\ref{defHNA}) with $n_f=5$, and evolve the operators down to a scale $\mu_b\sim m_b$ using Eqs.~(\ref{Unu}), (\ref{Upm}) and (\ref{Upmnu}). We decouple the $b$ quark using Eqs.~(\ref{Qnubdec}) and (\ref{ANApmbdec}) and continue evolving the operators with $n_f=4$ down to a scale $\mu_c\sim m_c$, where we express everything in terms of $Q_{\nu}$ using Eq.~(\ref{ANAcmatching}), and then find the expression of $X^{\rm{NA}}$.
We define the masses $\widetilde{m}_q \equiv m_q(m_q)$ and the logs $\ell_{\mu_q/\widetilde{m}_q} \equiv \ln(\mu_q/\widetilde{m}_q)$.
Further, we define the factors
\begin{align}
\label{defsKL}
K_c=\Bigg(\frac{a^{(4)}(\mu_c)}{a^{(4)}(\widetilde{m}_c)}\Bigg)\,,\quad
L=\Bigg(\frac{a^{(4)}(\mu_c)}{a^{(4)}(\mu_b)}\Bigg)^{\frac{1}{25}}\,,\quad 
M=\Bigg(\frac{a^{(5)}(\mu_b)}{a^{(5)}(\mu_W)}\Bigg)^{\frac{1}{23}}\,.
\end{align}
$K_c$  comes from expressing $m_{c}(\mu)$ in terms of $\widetilde{m}_{c}$, and $M$ and $L$ from the evolution from $\mu_W$ to $\mu_b$ and from $\mu_b$ to $\mu_c$, respectively.
With these conventions, our result reads
\begin{align}
\label{XNA}
X^{\rm{NA}}&=\frac{\widetilde{m}_c^2}{M_W^2}\frac{1}{32\,a^{(4)}(\mu_c)}\,K_c^{\frac{24}{25}}
 \Bigg\{ \tfrac{48}{7}\,L^{-6} M^{-6}+\tfrac{24}{11}\, L^{12} M^{12}\nonumber\\
 &+ L \big(- \tfrac{744}{65}\,M^{-1}+\tfrac{96}{35}\,M^{-6}  - \tfrac{48}{143}\, M^{12} \big)\nonumber\\
&+ a^{(5)}(\mu_W) \Big[ \tfrac{51784}{3703} L^{-6} M^{-6}- \tfrac{74968}{5819} L^{12} M^{12} + 
     L \big( \tfrac{2888594}{103155} M^{-1} + \tfrac{103568}{18515} M^{-6} + \tfrac{149936}{75647} M^{12} \big)\nonumber\\
&\hspace{1.6cm}+ \ell_{\mu_W/M_W} \Big( \tfrac{96}{7} L^{-6} M^{-6} - \tfrac{96}{11} L^{12} M^{12} + 
        L \big(-\tfrac{248}{65} M^{-1} + \tfrac{192}{35} M^{-6} + \tfrac{192}{143} M^{12}\big)\Big)\Big]\nonumber\\
&+ a^{(5)}(\mu_b) \Big[ -\tfrac{2609808}{2314375} L^{-6} M^{-6} + \tfrac{3769008}{3636875} L^{12} M^{12} + 
     L \Big( \tfrac{317527296}{21490625} M^{-1}- \tfrac{598688792}{34715625} M^{-6}\nonumber\\
&\hspace{1.6cm} - \tfrac{212453104}{141838125} M^{12}+ \ell_{\mu_b/\widetilde{m}_b} \big( \tfrac{496}{65} M^{-1}
- \tfrac{32}{5} M^{-6}- \tfrac{16}{13} M^{12}\big) \Big)\Big] \nonumber\\
&+ a^{(4)}(\mu_c) \Big[-\tfrac{966244}{13125} L^{-6} M^{-6} - \tfrac{57302}{6875} L^{12} M^{12}
+ \ell_{\mu_c/\widetilde{m}_c} \big(-16 L^{-6} M^{-6} + 8 L^{12} M^{12}\big)\nonumber\\
&\hspace{1.6cm}+  L \big( \tfrac{3772576}{40625} M^{-1}- \tfrac{486784}{21875} M^{-6}+ \tfrac{243392}{89375} M^{12}\big) \Big] 
+\mathcal{O}(a^2)\Bigg\}\,.
\end{align}
The $\mathcal{O}(a^2)$ coefficient is simply too long to print here. The full result can be retrieved from
\texttt{\bf http://www-ttp.particle.uni-karlsruhe.de/Progdata/ttp12/ttp12-048/}.
We will give numerical results for $X^{\rm{NA}}$ in section \ref{numres}.
The expression above agrees with the result of the NLO calculation performed in Refs.~\cite{Buchalla:1993wq,Buchalla:1998ba}, correcting for the fact that in those publications no decoupling was performed at $\mu_b$, and more importantly, that $\ln\left[\mu_c/m_c(\mu_c)\right]$ was simply changed by hand to $\ln\left[\mu_c/m_c(m_c)\right]$, and so they are missing a couple terms related to $K_c$.
Unfortunately, we could not  readily compare the final expression of our full result to that of Ref.~\cite{Buras:2006gb}, but we found agreement in the decoupling coefficients and anomalous dimensions leading to it (accounting in some cases for differences in normalization).

If we re-expand Eq.~(\ref{XNA}) around $\mu_W$ we obtain
\begin{align}
\label{XNAexp}
X^{\rm{NA}}&=\frac{\widetilde{m}_c^2}{M_W^2}\Bigg\{ \tfrac{1}{4}\left( \ell_{\widetilde{m}_c/M_W}+3\right) + 2\,a^{(5)}(\mu_W)\Big[\ell_{\mu_W/\mu_b}^2+\ell_{\mu_W/M_W}^2  + \ell_{\mu_b/\mu_c}\left(\tfrac{1}{6}-\ell_{\widetilde{m}_c/M_W}\right) \nonumber\\
&\hspace{1.8cm} + \ell_{\mu_b/\widetilde{m}_c}\left( \ell_{\mu_c/\widetilde{m}_c}-\ell_{\mu_W/M_W}-4\right) + \ell_{\mu_W/\mu_b}\Big(\ell_{\mu_b/\widetilde{m}_c}+\ell_{\mu_c/\widetilde{m}_c}-2\ell_{\mu_W/M_W}-\tfrac{23}{6}\Big)\nonumber\\
&\hspace{1.8cm} - \ell_{\mu_W/M_W}\left(  \ell_{\mu_c/\widetilde{m}_c} - \tfrac{23}{6} \right)
+ \tfrac{1}{6}\left(\ell_{\mu_c/\widetilde{m}_c} + 2\pi^2+29\right)\Big]\nonumber\\
&\hspace{1.8cm} + \Big(a^{(5)}(\mu_W)\Big)^2\Big[ \tfrac{8}{3}\ell_{\mu_W/\mu_b}^3 + 3\,\ell_{\mu_b/\mu_c}^3 
+\tfrac{1}{3}\ell_{\mu_b/\widetilde{m}_c}^2\Big(  \ell_{\mu_c/\widetilde{m}_c}-\ell_{\mu_W/M_W}+20\Big) \nonumber\\
&\hspace{1.8cm}+ 9\,\ell_{\mu_b/\mu_c}^2\Big(\ell_{\mu_W/\mu_b}-\ell_{\mu_W/M_W}+\ell_{\mu_c/\widetilde{m}_c}+\tfrac{5}{18}\Big) 
+\dots\Big]+\mathcal{O}\left(a^2\ell^0\right)\Bigg\}\,,
\end{align}
where for brevity's sake we only show a few terms of the long, last-order coefficient. If one differentiates this re-expanded expression with respect to the various $\mu_x$ scales present in it one finds a residual logarithmic dependence on them at the order $a^2$, which was of course not present in the original, perturbative calculation. The reason is that our treatment is only valid up to constant terms times $a^2$. Small logs like $\ell_{\mu_c/\widetilde{m}_c}$, $\ell_{\mu_b/\widetilde{m}_b}$, and $\ell_{\mu_W/M_W}$ are effectively taken as constants in our treatment, and thus beyond our reach in the last order of Eq.~(\ref{XNAexp}). By adding logs like these to Eq.~(\ref{XNAexp}) one can cancel its $\mu_x$ dependence, which serves as a check of the correctness of our result.

\section{Top quark anomalous diagram}
\label{secRGtop}

\subsection{Decoupling at the top scale}

\begin{figure}[t]
\centering
\subfloat[]{
 \includegraphics[height=.2\linewidth]{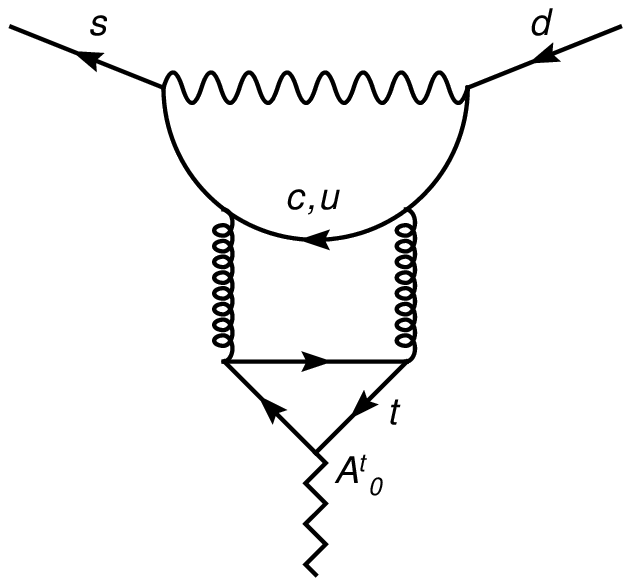}
\label{topanomalous}
}
\hspace{1cm}
\subfloat[]{
\includegraphics[height=.2\linewidth]{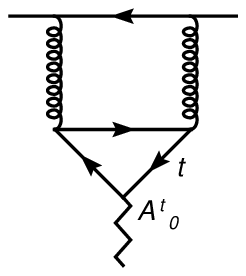}
\label{topdec}
}
\hspace{1cm}
\subfloat[]{
 \includegraphics[height=.2\linewidth]{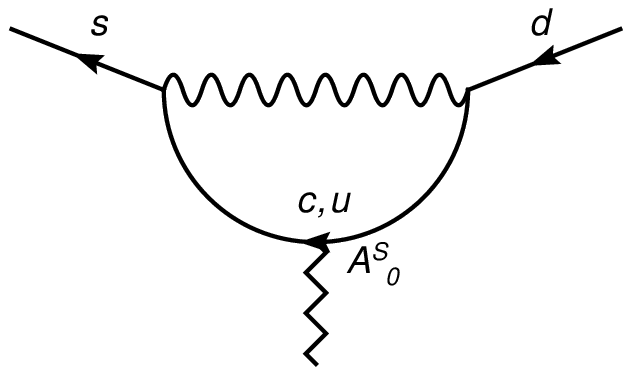}
\label{singletpenguin}
}
\caption{Top anomalous diagram (\ref{topanomalous}), with heavy subdiagram leading to $C_t$ (\ref{topdec}) and the remaining subdiagram (\ref{singletpenguin}).}
\end{figure}

The anomalous penguin diagram with a $W$ boson and a top quark is shown in Figure~\ref{topanomalous}, and the counterterm diagrams in Figure~\ref{pictcount}. Diagrams with the gluons attached to either one or both external quark lines vanish at out order ($1/m_t^0$). Much like in the non-anomalous case, when considering the decoupling of the top quark we have two distinct cases. The first one is when the loop momenta of the size of $m_t$ run only through the subdiagram in Figure~\ref{topdec}. The decoupling equation for this subdiagram reads
\begin{align}
 \big[A_{t}\big]^{(6)}=C_t\,\big[A^{\rm{S}}\big]^{(5)}~,\quad \big[A^{\rm S}\big]^{(5)}=\big[A^b+A^c+A^s+A^u+A^d\big]^{(5)} \label{Atdec}
\end{align}
The decoupling coefficient reads \cite{Chetyrkin:1993jm} ( $\ell_{\mu_t/m_t}=\ln[\mu_t^2/m_t^2(\mu_t)]$ )
\begin{align}
\label{decCt}
 C_t(\mu_t)=&~\Big(a^{(6)}(\mu_t)\Big)^2 \Big[2 - 8\,\ell_{\mu_t/m_t}\Big]~.
\end{align}
The resulting penguin diagram, with the operator $A^{\rm{S}}$ inserted, is shown in Figure~\ref{singletpenguin}. The second case is when all loop momenta are of the size of $m_t$. Then the whole diagram \ref{topanomalous} is treated as heavy and we get some coefficient $C_\nu$ times the local operator $Q_\nu$. As it turns out this coefficient is zero, i.e. the subdiagram shown in Figure \ref{topdec} factorizes from the penguin diagram.


\subsection{Decoupling at the $W$ scale}

When decoupling the $W$ we only have to treat the singlet penguin diagram in Figure \ref{singletpenguin}, which leads then, analogously  to the non-anomalous problem, to the effective Lagrangian
\begin{align}
 \mathcal{H}^{t}_{\nu,\rm{eff}} = C_t\,\big(C_+\,A^{\rm S}_+ + C_-\,A^{\rm S}_- + C^{\rm S}_\nu\,Q_\nu\big)~,
\end{align}
with the operator 
\begin{align}
 A^{\rm S}_{\pm}=-i\,\int \mathrm{d}^4x~\mathcal{T}\Big\{ \big(Q^c_{\pm}(x)-Q^u_{\pm}(x)\big) \,A^{\rm S}(0)\Big\}
\end{align}
and the coefficient ( $\ell_{\mu_W/M_W}=\ln[\mu_W^2/M_W^2]$ )
\begin{align}
 C^{\rm S}_\nu(\mu_W)=&~-a^{(5)}(\mu_W) \, 8\,(3+\ell_{\mu_W/M_W})~.
\end{align}

\subsection{Decoupling at the bottom and charm scales}

Because of the overall multiplying factor $C_t$, when decoupling the bottom quark we can neglect all terms of higher order than $a^2$.
This makes the decoupling equation quite trivial,
\begin{align}
\Big[A^{\rm S}_{\pm}\Big]^{(5)}=\Big[A^{\rm S}_{\pm}\Big]^{(4)} + \mathcal{O}\big(a^2\big)~.
\end{align}
In the next section, when we treat the bottom anomalous diagrams, we will give this decoupling equation to a higher order. The singlet current in the five flavor theory contains the bottom quark, while in the four flavor theory only the four lightest quarks are present. The decoupling equation for $Q_\nu$ is given in Eq.~(\ref{Qnubdec}).

At the charm scale, the matching equation reads
\begin{align}
\Big[A^{\rm S}_{\pm}\Big]^{(4)}=C^{\rm S}_{\pm,\nu}\,\Big[Q_{\nu}\Big]^{(4)}~. \label{AScmatching}
\end{align}
At our order, this matching is influenced only by non-anomalous diagrams generated by the $A^c$ and $A^u$ currents.
We got rid of non-anomalous contributions from $A^s$ and $A^d$ with our off-diagonal renormalization scheme, so these operators could only enter through anomalous diagrams. At the charm scale, however, these diagrams would only produce terms beyond our precision. 
Nevertheless, the matching for $A^{\rm S}_{\pm}$ at this scale will not the same as the one for $A^{\rm NA}_{\pm}$.
Since $A^{\rm NA}$ only appeared in open fermion lines we used the naive definition of $\gamma_5$ for it, but $A^{\rm S}$ appears here through the decoupling of the current $A^t$, which is inserted in a closed triangle loop. We used the definition in Eq.~(\ref{defAi}) for $A^t$, and use it too for $A^{\rm S}$. As mentioned in section \ref{secRGNA}, the different definitions of the axial-vector current produce different results for the penguin diagrams, which in turn leads to different constant parts in $C^{\rm S}_{\pm,\nu}$ compared to the coefficients $C^{\rm NA}_{\pm,\nu}$ given in Eqs.~(\ref{CNA+}) and (\ref{CNA-}), 
\newline( $\ell_{\mu_c/m_c}=\ln[\mu_c^2/m_c^2(\mu_c)]$  )
\begin{align}
 C^{\rm S}_{+,\nu}(\mu_c)=&16\,a^{(4)}(\mu_c) \Big[2 + \ell_{\mu_c/m_c}\Big] + \Big(a^{(4)}(\mu_c)\Big)^2 \Big[48 \ell_{\mu_c/m_c}^2 - 80 \ell_{\mu_c/m_c} - \tfrac{524}{3} \Big]~,\label{CAS+}\\
C^{\rm S}_{-,\nu}(\mu_c)=&-8\,a^{(4)}(\mu_c) \Big[2 + \ell_{\mu_c/m_c}\Big] - \frac{4}{3}\,\Big(a^{(4)}(\mu_c)\Big)^2 \Big[ 36 \ell_{\mu_c/m_c}^2+ 36 \ell_{\mu_c/m_c}+17\Big]~.\label{CAS-}
\end{align}
The minus sign between these and the previous expressions comes from the different  signs with which the currents $A^{c/u}$ enter in the definitions of $A^{\rm S}$ and $A^{\rm NA}$.

\subsection{Running to different scales}
\label{runS}

In contrast to $A^{\rm NA}$ the singlet operator $A^{\rm{S}}$ has a non-vanishing anomalous dimension
\begin{align}
\label{evS}
 \mu^2\frac{d}{d\mu^2}A^{\rm{S}}(\mu)=&~\gamma^{\rm{S}}\,A^{\rm{S}}(\mu)~,
\end{align}
with\cite{Chetyrkin:1993hk,Larin:1993tq}
\begin{align}
 \gamma^{\rm{S}}=-8\,n_f\,a^2 + a^3\,\Big[ -\frac{340}{3}\,n_f-\frac{8}{9}\,n_f^2\Big]~.
\end{align}
The solution of Eq.~(\ref{evS}) is
\begin{align}
 A^{\rm{S}}(\mu)
 =U_{\rm{S}}(\mu,\mu_0)\,A^{\rm{S}}(\mu_0)
 =\rm{exp}\bigg(\int_{a(\mu_0)}^{a(\mu)}\frac{dz}{z}\frac{\gamma_{\rm{S}}(z)}{\beta(z)}\bigg)\,A^{\rm{S}}(\mu_0)\,.\label{US}
\end{align}
For the bilocal operator we get the following RG equation,
\begin{align}
\mu^2\frac{d}{d\mu^2} A^{\rm S}_{\pm}(\mu)=\gamma^{\rm S}_{\pm}\, A^{\rm S}_{\pm}(\mu) + \gamma^{\rm S}_{\pm,\nu}\, Q_{\nu}(\mu)~, \label{evApmS}
\end{align}
with $\gamma^{\rm S}_{\pm}=\gamma_\pm + \gamma^{\rm S}$ and 
\begin{eqnarray}
\gamma^{\rm S}_{+,\nu}&=&16\,a-144\,a^2+a^3\Big[ -896\,\zeta_3 +\tfrac{800}{9}n_f+\tfrac{67682}{225} \Big]\,,\\
\gamma^{\rm S}_{-,\nu}&=&-8\,a+80\,a^2+a^3\Big[ 64\,\zeta_3+\tfrac{560}{9}n_f+\tfrac{147614}{225} \Big]\,.
\end{eqnarray}
The solution of Eq.~(\ref{evApmS}) is
\begin{align}
 \label{solveApmS}
 A^{\rm S}_{\pm}(\mu)= U^{\rm S}_{\pm}(\mu,\mu_0)\,A^{\rm S}_{\pm}(\mu_0) + U^{\rm S}_{\pm,\nu}(\mu,\mu_0)\,Q_\nu(\mu_0)~,
\end{align}
with $U^{\rm S}_{\pm}$ and $U^{\rm S}_{\pm,\nu}$ defined as in Eqs.~(\ref{Upm}) and (\ref{Upmnu}) with $\gamma_\pm$ ($\gamma^{\rm NA}_{\pm,\nu}$) replaced by $\gamma^{\rm S}_\pm$ ($\gamma^{\rm S}_{\pm,\nu}$).

\subsection{Result}

With the same definitions as in Eq.~(\ref{defsKL}), our result for the top anomalous diagrams reads
\begin{align}
X^{W,t}=&\frac{\widetilde{m}_c^2}{M_W^2}\frac{1}{32\,a^{(4)}(\mu_c)}\,K_c^{\frac{24}{25}}\cdot \Big( a^{(6)}(\mu_t)\Big)^2\Big[2 - 8\,\ell_{\mu_t/m_t}\Big]\bigg[\frac{48}{7}\,L^{-6} M^{-6}+\frac{24}{11}\, L^{12} M^{12} \nonumber\\
 &+ L \left(- \frac{744}{65}\,M^{-1}+\frac{96}{35}\,M^{-6}  - \frac{48}{143}\, M^{12} \right)\bigg]~.
\label{resWt}
\end{align}

\section{Bottom quark anomalous diagrams}
\label{secRGbottom}

\begin{figure}[t]
\centering
 \includegraphics[width=\linewidth]{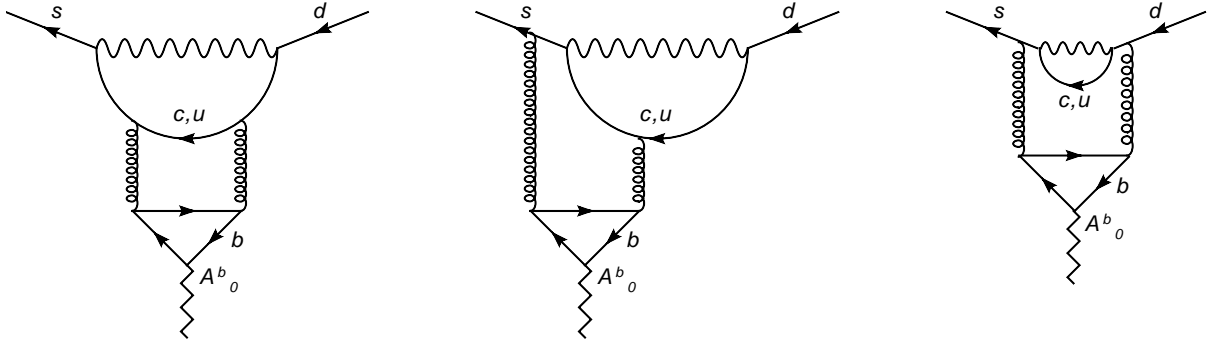}
\caption{The irreducible anomalous diagrams in the bottom case, omitting mirror images.\label{bottomanomalous}}
\end{figure}

The irreducible anomalous penguin diagrams with a $W$ boson and a bottom quark are shown in Figure~\ref{bottomanomalous}, and the counterterm diagrams in Figure~\ref{pictcount}.
At the order we are interested in there is no top quark appearing in the anomalous bottom diagrams. However, there are some considerations to be made before we start right away with the decoupling of the $W$ boson. As mentioned in section~\ref{secpert}, the explicit scale dependence of top and bottom anomalous diagrams cancels only in their sum. Since we started the evolution of top diagrams at the scale $\mu_t$, we must choose the same starting point in the bottom case, lest we end up with an unbalanced $\mu_t$ dependence in the final result. Alternatively, given that $\ell_{m_t/M_W}\sim 1.4$, we could simply perform the decoupling of the top quark at $\mu_W$ in the previous section. However, proceeding in this way effectively leads to the same problem. The top and $W$ scales may be relatively large, and their difference relatively small, but these anomalous diagrams start already at order $a^2$, and unlike the diagrams with the $\phi$ boson they contain also big logs involving the $m_c$ scale. Seemingly small differences can (and do) have noticeable effects in this case. 
Take the result from Eq.~(\ref{resWt}), substituting $\mu_t$ by $\mu_W$. Its $\mu_W$ dependence is, with our precision, formally the exact opposite of that of the result from the anomalous bottom diagrams starting at $\mu_W$, which will be presented later in Eq.~(\ref{resWb}). However, numerically the dependence is about an order of magnitude bigger in (the modified) Eq.~(\ref{resWt}) than in the bottom case, which breaks the balance present in the original perturbative results. Starting from $\mu_t$ is formally equivalent but will lead to better-behaved final expressions.

Thus, we consider the anomalous bottom diagrams starting from $\mu_t$. At our order, the decoupling of the top quark is trivial,
\begin{align}
\Big[ A^b\Big]^{(6)} = \Big[ A^b\Big]^{(5)} + \mathcal{O}(a^3)\,.
\end{align}
Next we must evolve our expressions to the scale $\mu_W$. Since the current $A^b$ mixes with $A^{\rm S}$ under the RG evolution, we make use of the identity
\begin{align}
\label{eqAb}
 A^b=A^b-\frac{1}{n_f}\,A^{\rm S} + \frac{1}{n_f}\,A^{\rm S}= A^{\rm NS} + \frac{1}{n_f}\,A^{\rm S}~,
\end{align}
with $A^{\rm NS}=A^b-1/n_f A^{\rm S}$, to express $A^b$ through the non-mixing operators $A^{\rm NS}$ and $A^{\rm S}$. Our choice of renormalization scheme for the axial current, presented in section \ref{defLarin}, ensures that $A^{\rm NS}$ has a vanishing anomalous dimension. The evolution of $A^{\rm S}$ was shown in Eq.~(\ref{US}). Thus, we have that
\begin{align}
\Big[A^b(\mu_t)\Big]^{(5)}&=\Big[A^{\rm NS}(\mu_W)\Big]^{(5)} +\frac{1}{5}U_S(\mu_t,\mu_W)\Big[A^{\rm S}(\mu_W)\Big]^{(5)}\nonumber\\
&=\Big[A^b(\mu_W)\Big]^{(5)} + \Big(\frac{24}{23}\left(a^{(5)}(\mu_t)-a^{(5)}(\mu_W)\right)+\mathcal{O}(a^2)\Big)\Big[A^{\rm S}(\mu_W)\Big]^{(5)}\,.
\end{align}
The treatment of $A^{\rm S}$ is now the same as in the top quark case, and exchanging $C_t$ by the coefficient from the expansion of $U_S$, the result will be the same as in Eq.~(\ref{resWt}), with a sign coming from the different isospins of top and bottom quarks. 
This all leaves us with the operator $\Big[A^b(\mu_W)\Big]^{(5)}$, whose treatment we present in the following sections.

\subsection{Decoupling at the $W$ scale}

As in the other cases we look at the different possible heavy subdiagrams to obtain the operators and coefficients in the effective theory. The effective Hamiltonian reads
\begin{align}
 \mathcal{H}^{b}_{\nu,\rm{eff}} = C_+\,A^{b}_+ + C_-\,A^{b}_- + C^{b}_\nu\,Q_\nu~,
\end{align}
with
\begin{align}
 A^{b}_{\pm}=-i\,\int \mathrm{d}^4x~\mathcal{T}\Big\{ \big(Q^c_{\pm}(x)-Q^u_{\pm}(x)\big) \,A^{b}(0)\Big\}\,,
\end{align}
and ( $\ell_{\mu_W/M_W}=\ln[\mu_W^2/M_W^2]$ )
\begin{align}
 C^{b}_\nu(\mu_W)=&~\Big(a^{(5)}(\mu_W)\Big)^3 \, \frac{8}{3}\Big[55+4\,\pi^2+18\,\ell_{\mu_W/M_W}+12\,\ell_{\mu_W/M_W}^2\Big]~.
\end{align}
 We will apply the same replacement to the bilocal operator as in Eq.~(\ref{eqAb}). The resulting operators $A^{\rm NS}_\pm$ and $A^{\rm S}_\pm$ can now be evolved separately from the $W$ to the bottom scale.

\subsection{Decoupling at the bottom and charm scales}

For the local operators $Q^q_\pm$ and $Q_\nu$ the decoupling equations are given in the previous sections. For $A^b$ the decoupling equation reads 
\begin{align}
 \Big[A^{b}\Big]^{(5)}=&~ C_b\,\Big[A^{\rm S}\Big]^{(4)}~, 
\end{align}
with ( $\ell_{\mu_b/m_b}=\ln[\mu_b^2/m_b^2(\mu_b)]$ )
\begin{align}
 C_b(\mu_b)=&~\Big(a^{(5)}(\mu_b)\Big)^2 \Big[2 - 8\,\ell_{\mu_b/m_b}\Big]
\end{align}
in complete analogy to the top decoupling in (\ref{Atdec}). Unlike the top quark case, however, the contribution from treating the whole diagram as heavy is not zero. Thus, the decoupling equation for the bilocal operator $A^{b}_{\pm}$ reads
\begin{align}
 \Big[A^{b}_{\pm}\Big]^{(5)}= B_\pm\,C_b\,\Big[A^{\rm S}_{\pm}\Big]^{(4)} + B^{b}_{\pm,\nu}\,\Big[Q_{\nu}\Big]^{(4)}~.\label{Abpmbdec}
\end{align}
where the coefficients $B_\pm$ have been defined in Eq.~(\ref{Qpmbdec}), and $B^{b}_{\pm,\nu}$ read
\begin{align}
 B^{b}_{+,\nu}(\mu_b)=&~ -\Big(a^{(5)}(\mu_b)\Big)^3 \Big[64\,\ell_{\mu_b/m_b}^2 + 64\,\ell_{\mu_b/m_b} + \tfrac{368}{3}   \Big]~,
 \\
 B^{b}_{-,\nu} (\mu_b) =&~ \Big(a^{(5)}(\mu_b)\Big)^3 \Big[32\,\ell_{\mu_b/m_b}^2 - 64\,\ell_{\mu_b/m_b} + \tfrac{328}{3} \Big]~.
\end{align}
At the order we are interested in, the operator $A^q$ for the four lightest quarks decouples naively ($[A^q]^{(5)}=[A^{q}]^{(4)}+\mathcal{O}(a^3)$). Also, as in the previous section, $A^s$ and $A^d$ make no contribution, since in our renormalization scheme they could only enter in anomalous diagrams that would lie beyond our precision. 
The decoupling of the bottom quark in a bilocal operator with an insertion of the current $A^c+A^u$ is given by
\begin{align}
 \Big[A^{(c+u)}_{\pm}\Big]^{(5)}&= B_\pm\,\Big[A^{(c+u)}_{\pm}\Big]^{(4)} + B^{(c+u)}_{\pm,\nu}\,\Big[Q_{\nu}\Big]^{(4)}~,
\label{ASpmbdec}
\end{align}
with
\begin{align}
 B^{(c+u)}_{+,\nu} (\mu_b) =&-\frac{8}{9}\,\Big(a^{(5)}(\mu_b)\Big)^3 \Big[ 12\,\ell_{\mu_b/m_b}^3+12\,\ell_{\mu_b/m_b}^2 + 107\,\ell_{\mu_b/m_b}+96\,\zeta_3-\tfrac{3616}{27}\Big]~,
 \\
 B^{(c+u)}_{-,\nu,u} (\mu_b) =&\frac{8}{9}\,\Big(a^{(5)}(\mu_b)\Big)^3 \Big[ 12\,\ell_{\mu_b/m_b}^3+12\,\ell_{\mu_b/m_b}^2 + 95\,\ell_{\mu_b/m_b}+96\,\zeta_3-\tfrac{3020}{27}\Big]~.
\end{align}
At our order, the difference between $B^{(c+u)}_{\pm,\nu}$ and $B^{\rm NA}_{\pm,\nu}$, besides an obvious sign, stems from the fact that here we are using the 't Hooft-Veltman definition of $\gamma_5$ instead of the naive one that we used for the non-anomalous diagrams, as mentioned before.
Using Eqs.~(\ref{Abpmbdec}) and (\ref{ASpmbdec}), and remembering the definition $A^{\rm NS}=A^b-1/5 A^{\rm S}$ we find at last that
\begin{align}
 \Big[A^{\rm NS}_{\pm}\Big]^{(5)}=&~ B_\pm\,\bigg( \frac{4}{5}\,C_b-\frac{1}{5}\bigg)\Big[A^{\rm S}_{\pm}\Big]^{(4)} 
       + \bigg( \frac{4}{5}\,B^{b}_{\pm,\nu}-\frac{1}{5}\,B^{(c+u)}_{\pm,\nu} \bigg)\,\Big[Q_{\nu}\Big]^{(4)}+\mathcal{O}\big(a^3\big)~,
 \\
 \Big[A^{\rm S}_{\pm}\Big]^{(5)}=&~ B_\pm\,\Big( C_b + 1 \Big)\Big[A^{\rm S}_{\pm}\Big]^{(4)} 
       + \Big( B^{b}_{\pm,\nu}+\,B^{(c+u)}_{\pm,\nu} \Big)\,\Big[Q_{\nu}\Big]^{(4)}+\mathcal{O}\big(a^3\big)~.
\end{align}
The matching at the charm scale is given in Eq.~(\ref{AScmatching}).


\subsection{Running to different scales}

The evolution of $A^{\rm{S}}_{\pm}$ was given in Eq.~(\ref{solveApmS}). $A^{\rm{NS}}$ has a vanishing anomalous dimension, and for $A^{\rm{NS}}_{\pm}$ we have
\begin{align}
\mu^2\frac{d}{d\mu^2} A^{\rm NS}_{\pm}(\mu)=\gamma_{\pm}\, A^{\rm NS}_{\pm}(\mu) + \frac{1}{5}\,\gamma_{\pm,\nu}^{\rm NS}\, Q_{\nu}(\mu)~, \label{evApmNS}
\end{align}
where
\begin{eqnarray}
\gamma_{+,\nu}^{\rm NS}&=&-16\,a+144\,a^2+a^3\Big[ 896\,\zeta_3+\tfrac{64}{9}n_f-\tfrac{67682}{225} \Big]\,,\\
\gamma_{-,\nu}^{\rm NS}&=&8\,a-80\,a^2-a^3\Big[ 64\,\zeta_3 +\tfrac{128}{9}n_f+\tfrac{147614}{225}\Big]\,.
\end{eqnarray}
The factor $1/5$ in Eq.~(\ref{evApmNS}) comes from the definition $A^{\rm NS}=A^b-1/5 A^{\rm S}$. 
The difference between $\gamma_{\pm,\nu}^{\rm NA}$, given in Eqs.~(\ref{anomdimNA+}) and (\ref{anomdimNA-}), and $\gamma_{\pm,\nu}^{\rm NS}$ comes again from the different definitions of $\gamma_5$.
The solution to Eq.~(\ref{evApmNS}) is
\begin{align}
\label{solveApmNS}
A^{\rm NS}_{\pm}(\mu)= U_{\pm}(\mu,\mu_0)\,A^{\rm NS}_{\pm}(\mu_0) + \frac{1}{5}\,U^{\rm NS}_{\pm,\nu}(\mu,\mu_0)\,Q_\nu(\mu_0)~,
\end{align}
with $U^{\rm NS}_{\pm,\nu}$ defined as in Eq.~(\ref{Upmnu}) with $\gamma_{\pm,\nu}^{\rm NA}$ replaced by $\gamma^{\rm NS}_{\pm,\nu}$.

\subsection{Result}

Again with the same definitions as in Eq.~(\ref{defsKL}), our result for the bottom anomalous diagrams reads
\begin{align}
 X^{W,b}=&~\frac{\widetilde{m}_c^2}{M_W^2}\frac{1}{32\,a^{(4)}(\mu_c)}\,K_c^{\frac{24}{25}}\bigg\{ 
 -\frac{24}{23}\left(a^{(5)}(\mu_t)-a^{(5)}(\mu_W)\right)\bigg[\frac{48}{7}\,L^{-6} M^{-6}+\frac{24}{11}\, L^{12} M^{12} \nonumber\\
 &+ L \left(- \frac{744}{65}\,M^{-1}+\frac{96}{35}\,M^{-6}  - \frac{48}{143}\, M^{12} \right)\bigg]
 +a^{(5)}(\mu_b)\bigg[ \tfrac{1152}{161}\,L^{-6}M^{-6} +\frac{576}{253}\,L^{12}M^{12} \nonumber\\
& - L\left(\frac{1600}{161}M^{-6} 
 +\frac{400}{253}\,M^{12}\right)\Big] +a^{(5)}(\mu_W)\bigg[-\frac{1152}{161}\,L^{-6}M^{-6}-\frac{576}{253}\,L^{12}M^{12}\nonumber\\
 &+L\left(\frac{912}{65}\,M^{-1}-\frac{2304}{805}\,M^{-6}+\frac{1152}{3289}\,M^{12}\right)
 \bigg]
  + \mathcal{O}(a^2)\bigg\}\,.\label{resWb}
\end{align}
The part proportional to $\left(a^{(5)}(\mu_t)-a^{(5)}(\mu_W)\right)$ comes from $\Big[A^{\rm S}(\mu_W)\Big]^{(5)}$, the rest from $\Big[A^b(\mu_W)\Big]^{(5)}$.
Like in the case of the non-anomalous diagrams, the $\mathcal{O}(a^2)$ term is too long to print here, and can be retrieved from
\\ \texttt {\bf http://www-ttp.particle.uni-karlsruhe.de/Progdata/ttp12/ttp12-048/}. 
There we also present a \texttt{Mathematica} file where all our results in the anomalous top and bottom cases are computed.
We give numerical results for $X^{W,b}$ in the following section.

In the original perturbative calculation the scale dependence of the sum $X^{W,t}+X^{W,b}$ is of order $a^3$. The sum of the resummed results of Eqs.~(\ref{resWt}) and (\ref{resWb}) satisfies this as well. However, as in the non-anomalous case, the re-expanded expression has some residual scale dependence at order $a^2$. It reads
\begin{align}
X^{W,t}+X^{W,b}=&\frac{\widetilde{m}_c^2}{M_W^2}\Big(a^{(6)}(\mu_t)\Big)^2
\Big[ 2\, \ell_{\mu_b/\mu_c} (  \ell_{\mu_W/\mu_t}+\ell_{\mu_t/\widetilde{m}_t} - \ell_{\mu_W/\widetilde{m}_b}  )
- \ell_{\mu_W/\mu_b}^2
 \nonumber\\
&\hspace{3.2cm} +2\,\ell_{\mu_W/\mu_b}  (\ell_{\mu_W/\mu_t} + \ell_{\mu_t/\widetilde{m}_t} - \ell_{\mu_c/\widetilde{m}_c})
\nonumber\\
&\hspace{3.2cm}- 2\,\ell_{\mu_W/\mu_t}  (1+\ell_{\mu_W/M_W} - \ell_{\mu_c/\widetilde{m}_c} )  
+\mathcal{O}\left(a^2\ell^0\right) \Big]~.\label{resWtWbexp}
\end{align}
The argument is the same as before. Our expansion misses constant terms times $a^2$, and small logs like $\ell_{\mu_c/\widetilde{m}_c}$, $\ell_{\mu_b/\widetilde{m}_b}$, $\ell_{\mu_W/M_W}$, and $\ell_{\mu_t/\widetilde{m}_t}$ are effectively constants here. Adding terms involving these logs to Eq.~(\ref{resWtWbexp}) one can cancel its scale dependence and bring it to essentially the same form (up to pure constants) as the purely perturbative result given by the sum of Eqs.~(\ref{pert1}) and (\ref{pert2}).

\section{Numerical Results and Discussion}
\label{numres}

In this section we present our numerical results for the non-anomalous and anomalous diagrams. The analytic expression for the former is given in Eq.~(\ref{XNA}), and those of the latter in Eqs.~(\ref{pert3}), (\ref{resWt}), and (\ref{resWb}). 
 We use the numerical solution of the differential equation relating $\alpha_s(\mu_0)$ with $\alpha_s(\mu)$ at two loops, together with the decoupling relations at threshold, which we take from the \texttt{Mathematica} package \texttt{RunDec}\cite {Chetyrkin:2000yt}.
We choose the following parameter  values ( $\widetilde{m}_q \equiv m_q(m_q)$ ),
\begin{equation}
\alpha_s(M_Z)=0.118\,,\nonumber
\end{equation}
\begin{equation}
\begin{array}{lllllllllllll}
M_W&=&80.4\,\text{GeV}\,,&\mu_W&=&M_W\,,\\
\widetilde{m}_t&=&162\,\text{GeV}\,,&\mu_t&=&\widetilde{m}_t\,,\\
\widetilde{m}_b&=&4.16\,\text{GeV}\,,&\mu_b&=&10\,\text{GeV}\,,\\
\widetilde{m}_c&=&1.29\,\text{GeV}\,,&\mu_c&=&3\,\text{GeV}\,.
\end{array}
\end{equation}

Our results are
\begin{itemize}
\item{Non-anomalous diagrams:
\begin{equation}
\label{nresXNA}
X^{\rm{NA}}=-(2\pm 2)\cdot 10^{-5}\,,
\end{equation}}
\item{Anomalous diagrams:
\begin{eqnarray}
X^{W,t}&=&-(1.8\pm 0.3)\cdot 10^{-8}\,,\label{nresXWt}\\
X^{W,b}&=&-(5\pm 5)\cdot 10^{-7}\,,\label{nresXWb}\\
X^{W,t+b}&=&-(5 \pm 5)\cdot 10^{-7}\,,\label{nresXWtb}\\
X^{\phi}&=&-(5\pm 2)\cdot 10^{-8}\,.\label{nresXphi}
\end{eqnarray}}
\end{itemize}
The uncertainty estimates were obtained by varying $\mu_W$ between $40$ and $160$ GeV, $\mu_b$ between $5$ and $15$ GeV and $\mu_c$ between $1.5$ and $4.5$ GeV. $X^{W,t+b}$ represents the sum of top and bottom anomalous diagrams, and in this case, in addition to the uncertainty estimates already mentioned, we added the one fom varying $\mu_t$ between $\widetilde{m}_t/2$ and  $2\widetilde{m}_t$ (we did not touch this scale in the separate top and bottom cases as only their sum should be independent of it).
The result for $X^{\phi}$ is simply the evaluation of Eq.~(\ref{pert3}), times $\pi^2\widetilde{m}_c^2/(2\,M_W^2\,g_s^2)$, evaluated at $\mu=M_W$. Its error, obtained by varying $\mu_W$ as described above, could be reduced with an RG treatment, but its size makes that treatment rather pointless. 

\begin{figure}[t]
\centering
\subfloat{
 \includegraphics[width=.49\linewidth]{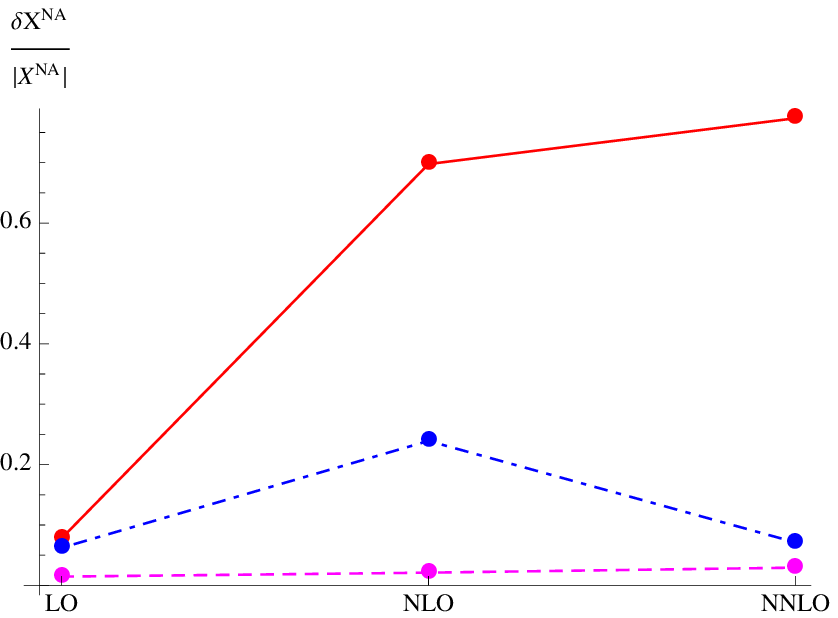}
 \label{XNApictures}}
 \subfloat{
 \includegraphics[width=.49\linewidth]{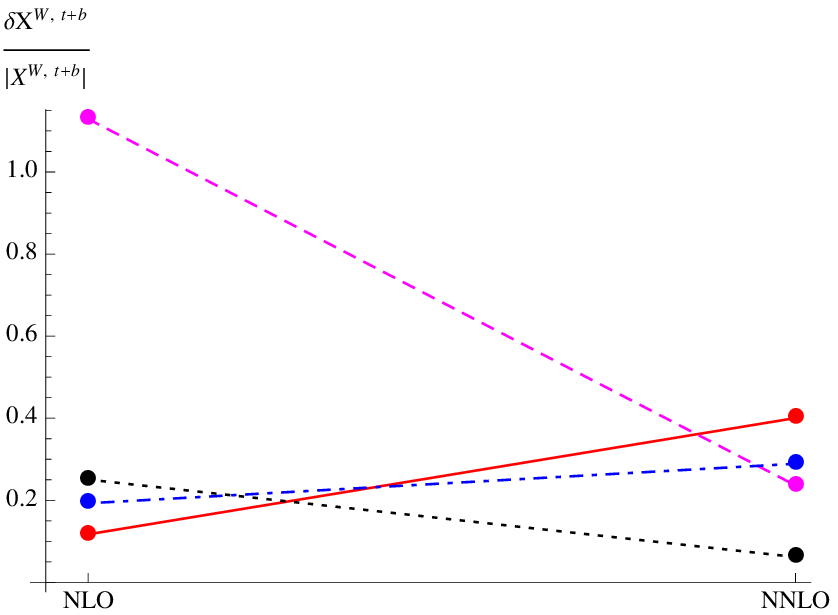}
 \label{XWbpictures}}
\caption{Evolution of the relative uncertainties at each order for $X^{\rm{NA}}$ and $X^{W,t+b}$. The bold line represents $\delta X_{\mu_c}$, the dashed one $\delta X_{\mu_b}$, the dot-dashed one $\delta X_{\mu_W}$, and the dotted one $\delta X_{\mu_t}$.\label{XNA-XWb}}
\end{figure}

We observe that the anomalous contributions, dominated by $X^{W,b}$, are two orders of magnitude smaller than the non-anomalous ones, and the uncertainty of the latter makes them completely negligible. Indeed, the uncertainties in for $X^{NA}$ and $X^{W,b}$ are of a remarkable size.
In Figure~\ref{XNA-XWb} we show the relative uncertainties of $X^{\rm{NA}}$ and $X^{W,t+b}$ at each order, that is $\delta X^{\rm{NA}}_{LO}/|X^{\rm{NA}}_{LO}|$, $\delta (X^{\rm{NA}}_{LO}+X^{\rm{NA}}_{NLO})/|X^{\rm{NA}}_{LO}+X^{\rm{NA}}_{NLO})|$ and so on. As we can see, $\delta X^{\rm{NA}}$ is completely dominated by the $\mu_c$ dependence, $\delta X^{\rm{NA}}_{\mu_c}$, which grows substantially in relative size at NLO, and keeps growing at NNLO. In the case of  $X^{W,t+b}$ the $\mu_c$ dependence still dominates and grows, although not to the same degree as in the non-anomalous case.

We observe uncertainties in the pictures growing when moving from one order to the next. This was unexpected, as normally uncertainties should diminish when increasing the order of our calculation. This odd increase in $\mu$ dependence, and the dominance of $\delta X_{\mu_c}$ especially in the non-anomalous case, seem to indicate that there might be a problem with the perturbation series, and that it might stem from the size of $\alpha_s(\mu_c)$. At $\mu=3$GeV, we have that $\alpha_s\simeq 0.252$.

\begin{figure}[hhh!]
\centering
\includegraphics[width=.49\linewidth]{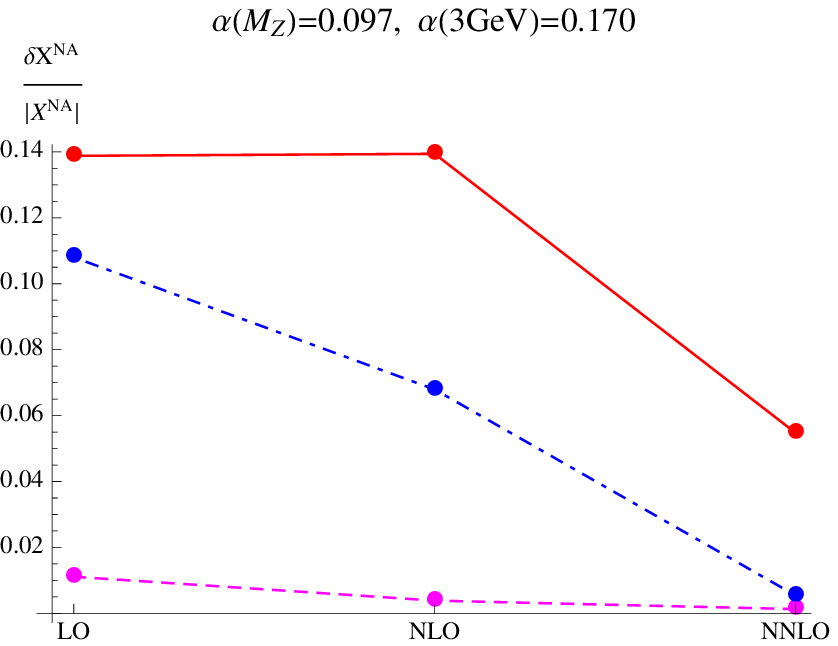}
 \includegraphics[width=.49\linewidth]{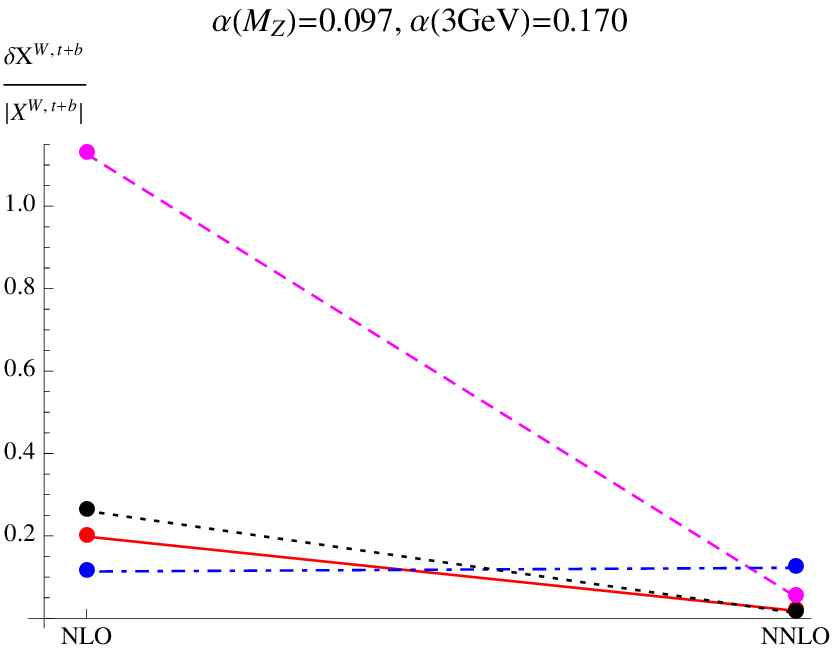}
\caption{Evolution of the relative uncertainties at each order for $X^{\rm{NA}}$ and $X^{W,t+b}$, with an artificially small $\alpha_s$. The bold line represents $\delta X_{\mu_c}$, the dashed one $\delta X_{\mu_b}$, the dot-dashed one $\delta X_{\mu_W}$, and the dotted one $\delta X_{\mu_t}$.\label{XNA2-XWb2}}
\end{figure}
\begin{figure}[hhh!]
\centering
\includegraphics[width=.49\linewidth]{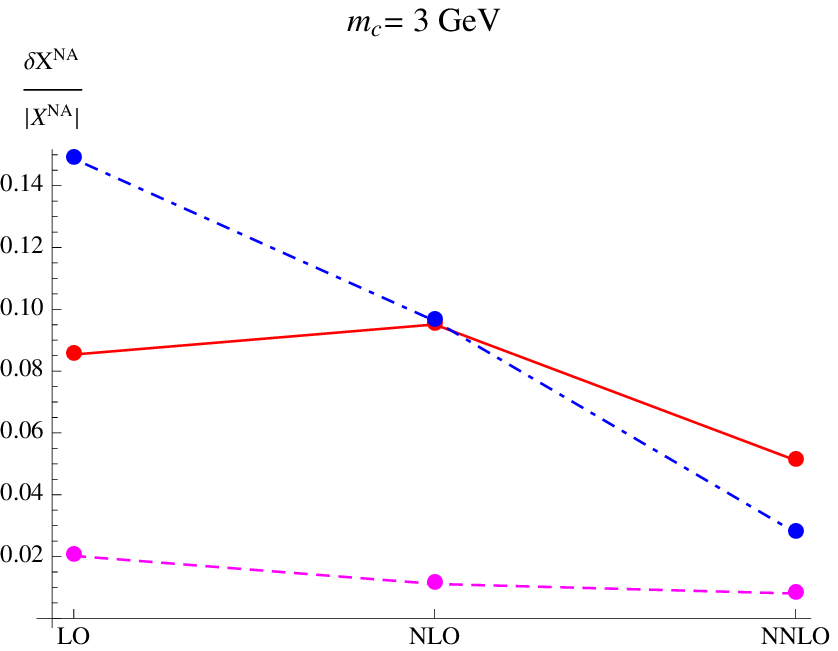}
 \includegraphics[width=.49\linewidth]{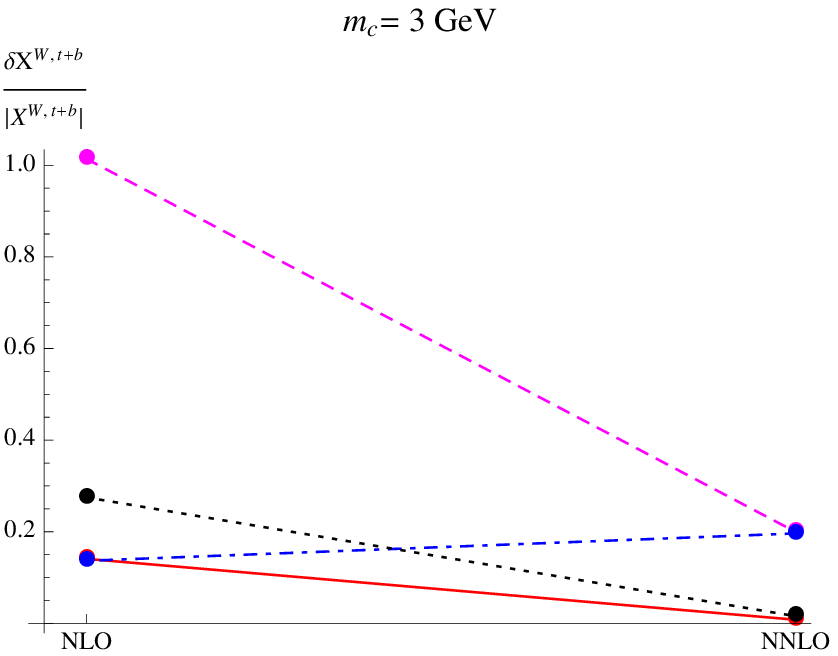}
\caption{Evolution of the relative uncertainties at each order for $X^{\rm{NA}}$ and $X^{W,t+b}$, setting $m_c=3$ GeV and $4.5$ GeV $<\mu_c<7.5$ GeV. The bold line represents $\delta X_{\mu_c}$, the dashed one $\delta X_{\mu_b}$, the dot-dashed one $\delta X_{\mu_W}$, and the dotted one $\delta X_{\mu_t}$\label{XNAm-XWbm}}
\end{figure}

We test this in two ways. First, we artificially set $\alpha_s(M_Z)=0.097$, which leads to $\alpha_s$(3GeV)$\simeq 0.170$. Second, we artificially increase the mass of the charm quark, setting it to $m_c=3$ GeV and varying $\mu_c$ in the range $4.5$ GeV$<\mu_c<7.5$ GeV. The results are shown in Figures~\ref{XNA2-XWb2} and \ref{XNAm-XWbm}.
When using the artificially smaller $\alpha_s$, the uncertainties in both $X^{\rm{NA}}$ and $X^{W,t+b}$ behave as we would have expected, except maybe for $\delta X^{W,t+b}_{\mu_W}$, which stays stable from one order to the next.
When using the heavier $m_c$, $\delta X^{\rm{NA}}_{\mu_c}$ still increases slightly at NLO, but its behavior is much milder than with the real $m_c$. In the case of the anomalous diagrams, $\delta X^{W,t+b}_{\mu_c}$ decreases and $\delta X^{W,t+b}_{\mu_W}$ increases slightly less than before.
From all this it seems that the low scale $\mu_c$ is really straining our perturbative treatment of the penguin diagrams, even with the log resummation provided by the RG evolution.

\subsection{Comparison with the previous calculation}
\label{comp}

As mentioned in section \ref{NAres}, the final analytic expression for $X^{\rm{NA}}$ could not be readily compared with that of Ref.~\cite{Buras:2006gb}, but we checked our agreement in the decoupling coefficients and anomalous dimensions that lead to it.
A direct comparison between our numerical results for $X^{\rm{NA}}$ is also not possible, since only the sum of the (non-anomalous) penguin and box contributions (also, they chose $\mu_c=1.5$ and $\mu_b=5$) was shown in Ref.~\cite{Buras:2006gb}. Following their treatment, we computed the box diagrams and found full agreement with the decoupling coefficients and the anomalous dimensions they presented. However, our calculation was incomplete, in that we did not include the effects of the mass of the $\tau$ lepton
at NNLO. Nevertheless, choosing their values for the parameters and the scales, and computing what they called the ``theory" error, which comes from varying the scales within the ranges they chose and adding the error from different treatments of $\alpha_s$ (which they stated
as $\pm0.001$ and we did not include in our results above), we obtained for the quantity $P_c(X)$ defined in their Eq.~(2) the result $P_c(X)=0.376\pm 0.01$, whereas in their Eq.~(119) they presented $P_c(X)=0.375+0.009_{theory}$. Thus, even with our incomplete box-diagrams calculation we have a very good agreement with their result.

Our value for $P_c(X)$ corresponds to $X=X^{\rm B}+X^{\rm{NA}}=(9.6\pm 0.3)\cdot 10^{-4}$. The relative error here looks much better than in all of the results in Eqs.~(\ref{nresXNA})-(\ref{nresXphi}), but this is simply because the result is dominated by the (incomplete) box contribution,
\begin{eqnarray}
X^{\rm B} = (9.7 \pm 0.2)\cdot 10^{-4}\nonumber\,,\\
X^{\rm{NA}}= -(1\pm 1)\cdot 10^{-5}\nonumber\,.
\end{eqnarray}
Our result for $X^{\rm{NA}}$ has changed from that of Eq.~(\ref{nresXNA}) because of the different choice of $\mu_b$ and $\mu_c$. 
It is worth noting that we did not observe any of the odd behavior of scale uncertainties with the box diagrams. It seems that only the penguin diagrams are so overly sensitive.

Unfortunately, in the anomalous case the situation is more complicated. In the Erratum of Ref.~\cite{Buras:2006gb}, the result for the anomalous diagrams is presented as $\Delta P_c(X)=1/\lambda^4(X^{W,t}+X^{W,b}+X^{\phi})$, where $\lambda=|V_{us}|$, for which they took the value $\lambda=0.2248$. As in the non-anomalous case, they chose the scales $\mu_c=1.5$ and $\mu_b=5$. Their treatment was again significantly different from ours, but unlike the non-anomalous case here not all the decoupling coefficients involved were shown.
However, they did show an expression of their analytic result after re-expansion, and a picture with the evolution of $\Delta P_c(X)$ as a function of $\mu_c$ in the range $1$~GeV~$<\mu_c<3$~GeV at NLO and NNLO. The result we get for $\Delta P_c(X)$ after re-expanding Eqs.~(\ref{resWt}) and (\ref{resWb}) and adding Eq.~(\ref{pert3}) only up to NNLO reads,
\begin{align}
\Delta P_c(X)=&\frac{1}{\lambda^4}\frac{\widetilde{m}_c^2}{M_W^2}\Big(\frac{\alpha_s(\mu_t)}{4\pi}\Big)^2
\Big[ 2\, \ell_{\mu_b/\mu_c} (  \ell_{\mu_W/\mu_t}+\ell_{\mu_t/\widetilde{m}_t} - \ell_{\mu_W/\widetilde{m}_b}  )
- \ell_{\mu_W/\mu_b}^2
 \nonumber\\
&\hspace{3.2cm} +2\,\ell_{\mu_W/\mu_b}  (\ell_{\mu_W/\mu_t} + \ell_{\mu_t/\widetilde{m}_t} - \ell_{\mu_c/\widetilde{m}_c})
\nonumber\\
&\hspace{3.2cm}- 2\,\ell_{\mu_W/\mu_t}  (1+\ell_{\mu_W/M_W} - \ell_{\mu_c/\widetilde{m}_c} )
\nonumber\\
&\hspace{3.2cm} -\tfrac{1}{4}\ell_{M_W/\widetilde{m}_t}^2 +\tfrac{3}{2}\ell_{M_W/\widetilde{m}_t}
+\mathcal{O}\left(\alpha_s^{2}\ell^{0}\right) \Big]~.\label{reexp}
\end{align}
As mentioned before, there is a scale dependence in this expression that comes simply from the fact that our expansion misses terms of order $\alpha_s^2$ times some small log, like $\ell_{\mu_x/\widetilde{m}_x}$. Such logs are effectively constants in our treatment and therefore missed at NNLO, and the fact that they are all that is needed at this order to cancel the scale dependence of Eq.~(\ref{reexp}) and bring it to essentially the same form as the perturbative result (up to constants) is a good check of our calculations. 

In Ref.~\cite{Buras:2006gb} the result for $\Delta P_c(X)$ is given as
\begin{align}
\label{reexpErr}
\Delta P_c(X)=&\frac{1}{\lambda^4}\frac{m_c^2}{M_W^2}\Big(\frac{\alpha_s}{4\pi}\Big)^2
\Bigg\{ \left( \ln\frac{\mu_W^2}{\mu_b^2} - 2\,\ln\frac{\mu_W^2}{\mu_c^2} +1\right)\ln\frac{\mu_W^2}{\mu_b^2}\Bigg\}+\mathcal{O}\left(\alpha_s^{2}\ln^{0}\frac{\mu_W^2}{\mu_b^2},\alpha_s^3\right)~.
\end{align}
This expression is significantly different from ours. The first thing one notices is the absence of logs of masses in it, which leads to the impossibility of canceling its scale dependence with small logs of the type $\ell_{\mu_x/\widetilde{m_x}}$. One also notices the absence of $\mu_t$, which makes us believe that the top and $W$ scales were considered equal. However, even if $\ell_{\mu_t/M_W}$ is considered small, in the perturbative result for $D^{W,t}$ in Eq.~(\ref{pert1}) we see that it is multiplied by a big log, namely $\ell_{m_c/M_W}$, and therefore one should have either $\mu_t$ or $m_t$ in the final result. Finally, in Eq.~(\ref{reexpErr}) there is a linear $\ell_{\mu_W/\mu_b}$ that is not present in our case. In any case, regardless of the discrepancies of our analytic expressions, the size of the uncertainties makes our numerical results and theirs basically the same, both negligible compared to the non-anomalous case anyway.
\begin{figure}[t]
\centering
\subfloat[]{
\includegraphics[width=.47\linewidth]{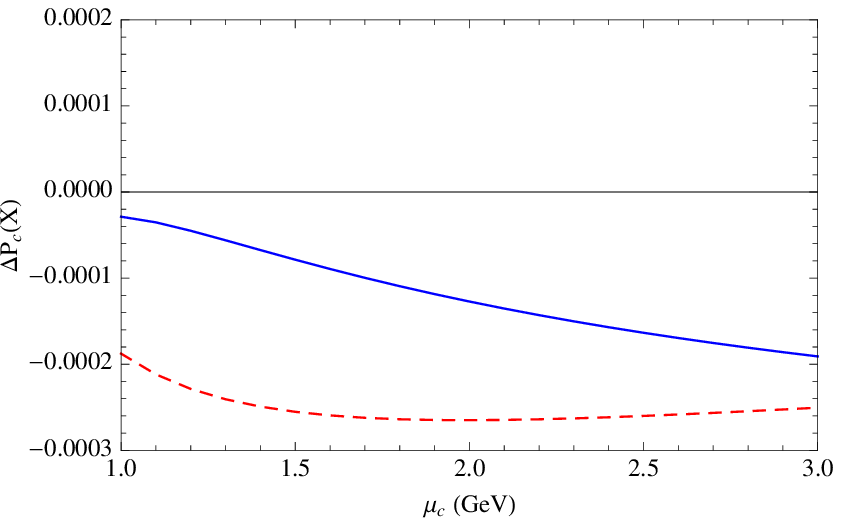}
}
\hspace{0.1cm}
\subfloat[]{
\includegraphics[width=.47\linewidth]{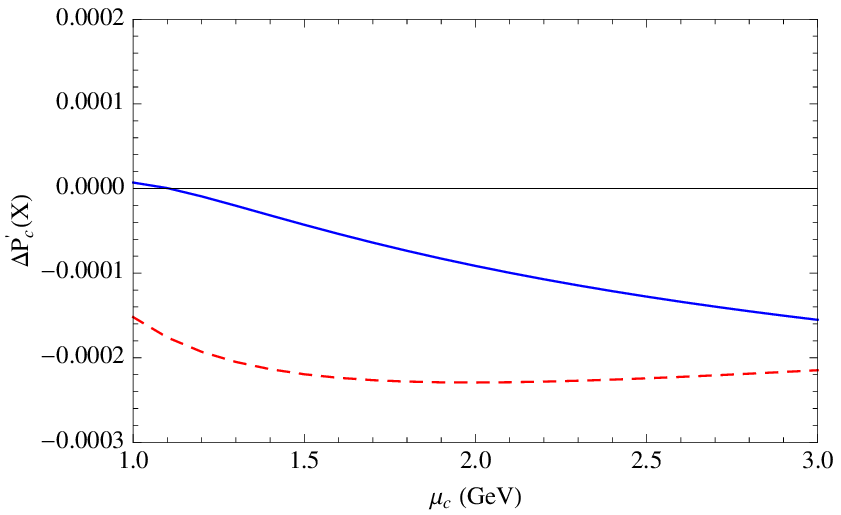}
}
\caption{$\Delta P_c(X)$ (a) and $\Delta P'_c(X)$ (b) as a function of $\mu_c$ in the range $1$ GeV $<\mu_c<3$ GeV at NLO (dashed line) and NNLO (solid line).\label{DeltaP}}
\end{figure}

Regarding the evolution of $\Delta P_c(X)$ with $\mu_c$ at different orders, given that $X^{\phi}$ is independent of this (or, to our order, any) scale, we simply add $X^{\phi}$ as a constant both at NLO and NNLO. Alternatively, we define $\Delta P'_c(X)=\Delta P_c(X-X^{\phi})$.
Our results are shown in Figure~\ref{DeltaP}. Neither plot agrees with that of Ref.~\cite{Buras:2006gb}, where the curves have the same shape but are translated upwards. 
Guided by the absence of the top scale in their re-expanded expression, we can find agreement with their picture if we plot $\Delta P'_c(X)$ and in the bottom case we neglect the running from $\mu_t$ to $\mu_W$, that is, if we compute the top case starting from $\mu_t$ and the bottom case starting from $\mu_W$. This produces essentially the same results within the errors, but as mentioned before it produces an unbalanced $\mu_t$ dependence which was not present in the original perturbative diagrams.



\section{Conclusion}
\label{conclusion}

In this paper we set out to check the results for penguin-type charm contributions to the decay $K^+\to \pi^+\nu\bar{\nu}$ presented in Ref.~ \cite{Buras:2006gb}. It is quite an involved calculation, certainly deserving an independent check. We have found full agreement for all the decoupling coefficients and anomalous dimensions for the non-anomalous diagrams. In Ref.~ \cite{Buras:2006gb} the final numerical result presented was the sum of non-anomalous penguin and box-type diagrams, so in order to perform a comparison we also had to compute the latter. Our calculation missed $\tau$-mass effects, but again we confirmed the corresponding decoupling coefficients and anomalous dimensions presented in Ref.~ \cite{Buras:2006gb}. Incomplete as our result for the box diagrams was, it was enough to allow us to obtain very good agreement in the final numerical result, both in the central value and the estimate of theoretical uncertainty.

We found some discrepancies with Ref.~ \cite{Buras:2006gb} in our results for the anomalous diagrams. Although numerically the differences get washed out by the uncertainty (and by the sheer smallness of the anomalous contributions versus the non-anomalous ones), the analytic expression of our re-expanded result differs significantly from that of Ref.~ \cite{Buras:2006gb}.

Our results show some unstable behavior and quite a large uncertainty. The cause seems to be the low scale $\mu_c$, which might be testing the limits of perturbation theory. This only applies to the penguin diagrams, however, as in our evaluation of the box diagrams the scale uncertainties behaved as expected, diminishing at each order and remaining relatively small. We were not able to find an analytic reason for the unique unstable behavior of the penguin diagrams. 

\medskip

{\bf Acknowledgments}.  We are grateful to K.G.~Chetyrkin for his invaluable help and support throughout this project.
We also wish to thank A.~Buras, M.~Gorbahn, U.~Haisch, and U.~Nierste for helpful discussions and for sharing their Erratum 
with us prior to publication. 
This work was supported by the Deutsche Forschungsgemeinschaft in the
Sonderforschungsbereich/Transregio SFB/TR9 ``Computational Particle Physics".

\appendix
\section{Definition of the evanescent operators}
\label{appe}

In dimensional regularization the operators $Q_{\pm}^q$ introduced in Eq.~(\ref{defQpm}) are not enough to perform the decoupling of the $W$ boson in the subdiagram shown in Figure~\ref{4qLO}. A set of evanescent operators (vanishing at $d=4$) is required. In our renormalization scheme these operators have vanishing matrix elements, but they still contribute by determining the anomalous dimensions of $Q_{\pm}^q$. Since we reach $\mathcal{O}(\alpha_s^2)$ in our calculations, we need six evanescent operators. We take the definitions from Ref.~\cite{Buras:2006gb}, with a different normalization factor, coming from using a $(V-A)$ current instead of left-handed fields. If we define
\begin{eqnarray}
Q_1^q&=& \big(\bar{s}\gamma^{\mu}(1-\gamma_5)t^aq\big)\big(\bar{q}\gamma_{\mu}(1-\gamma_5)t^ad\big)\\
Q_2^q&=& \big(\bar{s}\gamma^{\mu}(1-\gamma_5)q\big)\big(\bar{q}\gamma_{\mu}(1-\gamma_5)d\big)\,,
\end{eqnarray}
where $t^a$ is a generator of the color group, then one can define the following evanescent operators,
\begin{eqnarray}
E_1^q&=&\big(\bar{s}\gamma^{\mu_1\mu_2\mu_3}(1-\gamma_5)t^aq\big)\big(\bar{q}\gamma_{\mu_1\mu_2\mu_3}(1-\gamma_5)t^ad\big)
-(16-4\epsilon-4\epsilon^2)Q_1^q\,,\\
E_2^q&=&\big(\bar{s}\gamma^{\mu_1\mu_2\mu_3}(1-\gamma_5)q\big)\big(\bar{q}\gamma_{\mu_1\mu_2\mu_3}(1-\gamma_5)d\big)
-(16-4\epsilon-4\epsilon^2)Q_2^q\,,\\
E_3^q&=&\big(\bar{s}\gamma^{\mu_1\mu_2\mu_3\mu_4\mu_5}(1-\gamma_5)t^aq\big)\big(\bar{q}\gamma_{\mu_1\mu_2\mu_3\mu_4\mu_5}(1-\gamma_5)t^ad\big)
\nonumber\\
&&-(256-224\epsilon-\tfrac{5712}{25}\epsilon^2)Q_1^q\,,
\\
E_4^q&=&\big(\bar{s}\gamma^{\mu_1\mu_2\mu_3\mu_4\mu_5}(1-\gamma_5)q\big)\big(\bar{q}\gamma_{\mu_1\mu_2\mu_3\mu_4\mu_5}(1-\gamma_5)d\big)
\nonumber\\
&&-(256-224\epsilon-\tfrac{10032}{25}\epsilon^2)Q_2^q\,,
\\
E_5^q&=&\big(\bar{s}\gamma^{\mu_1\mu_2\mu_3\mu_4\mu_5\mu_6\mu_7}(1-\gamma_5)t^aq\big)\big(\bar{q}\gamma_{\mu_1\mu_2\mu_3\mu_4\mu_5\mu_6\mu_7}(1-\gamma_5)t^ad\big)
\nonumber\\
&&-(4096-7680\epsilon)Q_1^q\,,
\\
E_6^q&=&\big(\bar{s}\gamma^{\mu_1\mu_2\mu_3\mu_4\mu_5\mu_6\mu_7}(1-\gamma_5)q\big)\big(\bar{q}\gamma_{\mu_1\mu_2\mu_3\mu_4\mu_5\mu_6\mu_7}(1-\gamma_5)d\big)
\nonumber\\
&&-(4096-7680\epsilon)Q_2^q\,.
\end{eqnarray}
Here the notation $\gamma_{\mu_1\mu_2\dots\mu_n}=\gamma_{\mu_1}\gamma_{\mu_2}\dots\gamma_{\mu_n}$ has been used.
The operators $Q_{1,2}^q$ are combined to produce $Q_{\pm}^q$ in the following way,
\begin{eqnarray}
Q_+^q&=&Q_1+\frac{n_c+1}{2n_c}Q_2\,,\\
Q_-^q&=&-Q_1+\frac{n_c-1}{2n_c}Q_2\,,
\end{eqnarray}
where $n_c$ is the number of colors.
In the same way, the pairs $(E^q_3,E^q_4)$ and $(E^q_5,E^q_6)$ are combined to generate $E_{\pm}$ operators, which are the ones we use. The above definitions of the evanescent operators ensure that $Q_{\pm}^q$ do not mix with each other through NNLO.

\bibliographystyle{JHEP}

\bibliography{literatura}{}

\end{document}